\documentclass[journal,doublecolumn,10pt,romanappendices]{IEEEtran}

\usepackage{url}
\usepackage{ifthen}
\usepackage{graphics}
\usepackage{cite}
\usepackage{amssymb}
\usepackage{bm}
\usepackage{bbm}
\usepackage{xcolor}
\usepackage[cmex10]{amsmath} 

\usepackage{graphicx}
\usepackage{subfigure}
\usepackage[ruled,linesnumbered] {algorithm2e}
\newtheorem{lemma}{Lemma}
\newtheorem{theorem}{Theorem}

\newtheorem{definition}{Definition}

\begin{document}
\title{Degree-of-Freedom of Modulating Information in the Phases of Reconfigurable Intelligent Surface}

\author{
{Hei Victor Cheng}, \IEEEmembership{Member, IEEE},
        and Wei Yu, \IEEEmembership{Fellow, IEEE}
\thanks{Hei Victor Cheng was with The Edward S.\ Rogers Sr.\ Department of Electrical and Computer Engineering, University of Toronto, Canada. He is now with the Department of Electrical and Computer Engineering, Aarhus University, 8200 Aarhus, Denmark. Wei Yu is with The Edward S.\ Rogers Sr.\ Department of Electrical and Computer Engineering, University of Toronto, Toronto, ON M5S 3G4, Canada. (e-mails: hvc@ece.au.dk, weiyu@ece.utoronto.ca). The work has been presented in part at {\it IEEE International Symposium on Information Theory}, July 2021 \cite{dof_isit2021} and at {\it IEEE International Symposium on Wireless Communication Systems}, October 2022 \cite{cheng2022}. This work was supported by Huawei Technologies Canada Ltd. Co. and by the Natural Sciences and Engineering Research Council (NSERC) of Canada. The work of Hei Victor Cheng was supported in addition by the Aarhus Universitets Forskningsfond under Project AUFF 39001.
}
}

\maketitle
\begin{abstract}
This paper investigates the information theoretic limit of a reconfigurable
intelligent surface (RIS) aided communication scenario in which the RIS and the
transmitter either jointly or independently send information to the receiver.
The RIS is an emerging
technology that uses a large number of passive reflective elements with
adjustable phases to intelligently reflect the transmit signal to the intended
receiver.  While most previous studies of the RIS focus on its ability to
beamform and to boost the received signal-to-noise ratio (SNR), this paper
shows that if the information data stream is also available at the RIS and can
be modulated through the adjustable phases at the RIS, significant
improvement in the {degree-of-freedom} (DoF) of the overall channel is possible.
For example, for an RIS system in which the signals are reflected from a
transmitter with $M$ antennas to a receiver with $K$ antennas through an RIS
with $N$ reflective elements, assuming no direct path between the transmitter
and the receiver, joint transmission of the transmitter and the RIS can achieve
a DoF of $\min\left(M+\frac{N}{2}-\frac{1}{2},N,K\right)$ as compared to the DoF of
$\min(M,K)$ for the conventional multiple-input multiple-output (MIMO) channel.
This result is obtained by establishing a connection between the RIS system and
the MIMO channel with phase noise and by using results for characterizing the
information dimension under projection.  The result is further extended to the case
with a direct path between the transmitter and the receiver, and also to the
multiple access scenario, in which the transmitter and the RIS send independent
information.
Finally, this paper proposes a symbol-level precoding approach for
modulating data through the phases of the RIS, and provides numerical
simulation results to verify the theoretical DoF results.
\end{abstract}


\section{Introduction}

Reconfigurable intelligent surface (RIS), also known as the intelligent
reflective surface (IRS), is a promising new technology that utilizes a
large number of reflective elements with adjustable phase shifts to enhance
the spectrum and energy efficiencies of wireless communication systems
\cite{reflecting_metasurface2017,basar2019wireless,di2019smart,liang2019large}.
In most of the current literature, the RIS is envisioned to be used as a passive
beamformer with the reflective coefficients being configured for maximizing the
received signal-to-noise ratio (SNR) of the overall channel. Passive
beamforming, however, does not fully harvest the potentials of the RIS.
The goal of this paper is to show that if the information data stream is
available at the RIS, it is possible to use the RIS to modulate the information
through phase shifts. In effect, the adjustable reflective coefficients can
be used not only to enhance the channel, but also to carry information.

This paper aims to show that modulating information through phases
allows the RIS to significantly enhance the degree-of-freedom (DoF),
also known as the multiplexing gain, of the overall channel between the
transmitter and the receiver.  The gains in DoF from modulating information
through the phases of the RIS are quantified both for the cases with and
without cooperation between the transmitter and the RIS. The former case
corresponds to the point-to-point multiple-input multiple-output (MIMO) channel
where the RIS helps the transmitter send the same data, while the latter case
corresponds to a multiple access channel where the RIS and the transmitter
send independent data which are jointly decoded at the receiver.

Moreover, this paper proposes a symbol-level precoding strategy for the joint
transmission of information from the transmitter and the RIS to the receiver.
We show that such a strategy can approach the information theoretic
DoF of the overall system.

\subsection{Prior Works}

RIS can be thought of as an array of a large number of reflective elements, each of which can induce a phase shift between the incident signal and the reflecting signal.  By adjusting the phase shifts, the RIS can enhance the transmission quality of the overall channel by adaptively beamforming the incident signal to an intended reflecting direction \cite{wu2019intelligent}.  As compared to traditional relaying techniques
\cite{emil2020,ntontin2019reconfigurable}, the RIS reflects the incident signal passively, so it has a much lower energy consumption. Further, by utilizing a large number of analog reflective elements and by adjusting their phases in real time, the RIS can achieve a high beamforming gain at relatively low hardware cost.

Because of the above promising advantages, the applications of RIS for
enhancing wireless communication systems have been extensively investigated in
the literature. Most of the existing literature on the RIS-aided wireless
communication system focuses on the optimization of the phase shifts at the RIS
to improve various system-level objectives e.g., for energy minimization
\cite{huang2019reconfigurable}, maximizing the minimum
signal-to-interference-plus-noise ratio (SINR) \cite{huang2019reconfigurable},
for sum-rate maximization for both single-cell \cite{guo2019weighted} and
multi-cell \cite{pan2019multicell} scenarios, for utility maximization using a machine learning approach \cite{jiang2020}, as well as for various scenarios such as the wideband orthogonal frequency
division multiplexing (OFDM) system \cite{yang2019irs}, the millimeter-wave
(mmWave) system \cite{wang2019intelligent, perovic2019channel}, and the multiple RIS systems
\cite{jung2019optimality}.
These existing studies, however, utilize the RIS only as a passive analog
beamformer to achieve an SNR gain.
Passive analog beamforming works by adjusting the phase shifts of
scattered wavefronts at different reflective elements so that they
add constructively at the receiver, thereby enhancing the SNR at the receiver.
For these systems, the phase shifts at the RIS are functions of the
instantaneous channel realizations only.

The main focus of this paper is to investigate the possibility that if, in addition
to the channel state information (CSI), the RIS also has access to the data,
then the RIS can modulate the information stream in the phase shifts to further
enhance the capacity of the overall communication channel. By modulating
information using the RIS, the information is essentially being conveyed by
changing the transmission environment.

The idea of modulating information through the transmission environment
has been considered by several earlier works under different names, e.g.
media-based communication \cite{media_modulation} and load modulated arrays
\cite{load_modulated}.  But an exact analysis of the information theoretic
limits of such systems is a challenging task.  In the RIS context, this
idea has been previously explored in \cite{karasik2021}, which studies the
channel capacity for joint information transmission by the transmitter and the RIS, but under the assumption that the RIS operates with a fixed finite modulation and the transmitter is equipped with single antenna. The same analysis has been extended to the multiple access channel under the same setup in \cite{karasik2021isit} where the transmitter and the RIS are sending different messages.
Information transmission through RIS
has also been considered in \cite{yan2020} and \cite{reflecting_modulation},
which primarily focus on the design of decoding algorithms when extra
information bits are carried in the RIS. In general, the information
theoretic limit of modulating through RIS has not been explicitly characterized in computable form.
For example, the finite modulation assumption of \cite{karasik2021, karasik2021isit} does not lead to an analysis of the channel capacity at high SNR.



\subsection{Main Contributions}

This paper aims to characterize the DoF, or the multiplexing gain,
or the pre-log factor, of a wireless transmission system equipped with an RIS in
which information is modulated through the phase shifts. 
We study both the case
in which the information bits are jointly encoded in the transmit radio frequency
(RF) signal and the phase shifts at the RIS for an intended MIMO receiver,
as well as the case in which the RIS and the transmitter send independent
information, corresponding to a multiple access channel.



The DoF is an important metric that captures the high SNR asymptotic behavior of
the capacity for channels with additive noise. The DoF is the number of data
streams that a system can support at high SNR, which is an important consideration
for practical system design. 
The main challenge in the analysis of the
RIS channel lies in the multiplicative nature of the channel model in which the
received signal is the product of two different information-carrying signals.
We tackle this challenge by connecting the achievable rate of the RIS channel
to the capacity of the MIMO channel with phase noise. Tools for studying the
information dimension and the point-wise dimension under projection are used.

For an RIS system consisting of a transmitter with $M$ transmit antennas, a
receiver with $K$ receive antennas, and an RIS with $N$ reflective elements,
the main results of this paper show that:
\begin{itemize}
\item For the case of joint transmission by the transmitter and the RIS,
the DoF of the overall channel is $\min(M+\frac{N}{2}-\frac{1}{2}, N, K)$ 
, assuming that the direct path between the
transmitter and the receiver is blocked.
The $1/2$ factor corresponds to the phase ambiguity between the RF signal and the RIS phase shifts due to the multiplicative nature of the channel.

\item When there is a direct path with channel rank $r>0$, the DoF of joint
transmission is $\min(M+\frac{N}{2}, N+r, K)$.

This implies that the phase ambiguity can be resolved when a direct path is present.

\item The above results imply that as compared to a conventional MIMO channel
with $M$ transmit antennas and $K$ receive antennas achieving a DoF of
$\min(M,K)$, in the case of $M<K$, a cooperative RIS can significantly
improve the overall channel capacity by allowing data modulation through
the phase shifts.
\item When the transmitter and the RIS send independent information,
the same sum DoF can also be achieved as the above joint
transmission case. This means that if there are more receive antennas than
transmit antennas, the RIS can embed additional data streams in the phase
shifts without affecting the DoF of the direct transmission.
\item A symbol-level precoding strategy is proposed as a practical
implementation of jointly encoding information by the transmitter and the
RIS. It is shown to approach the DoF of joint
transmission.
\end{itemize}

\subsection{Paper Organization and Notation}
The rest of the paper is organized as follows.
Section \ref{sec:model} describes the channel model for the RIS system.
Section \ref{sec:no_direct_path} discusses the RIS model without a direct
path and provides the DoF result for both the joint transmission case
and the multiple access case.
These results are extended to the case in which the direct path is present
in Section \ref{sec:direct_path} and to the case in which the RIS is operating at a different symbol rate as the transmitter in Section \ref{sec:different_rate}. Practical implementation of the joint
transmission strategy via symbol-level precoding is discussed in
Section \ref{sec:slp}. Numerical results are presented in Section \ref{sec:sim}.
Conclusions are drawn in \ref{sec:conclusion}.

The notations used in the paper are as follows. Lower-case letters are used
to denote scalars; upper-case letters are used to denote random variables.
Bold-faced letters are vectors. Bold-faced upper-case letters are either random
vectors, or matrices, where the context should make it clear. Exceptions to the
above are the dimensions of the transmitter, the RIS, and the receiver vectors,
which are denoted as $M$, $N$, and $K$, respectively.
The capacity of a channel with power constraint $P$ is denoted as $C(P)$.
Finally, we use $\mathcal{CN}(\cdot,\cdot)$ to denote the circularly symmetric
complex Gaussian distribution
and $\operatorname{Unif}(\cdot)$ to denote the uniform distribution.

\section{System Model and Problem Formulation}
\label{sec:model}

\subsection{Channel Model}

\begin{figure}
        \centering
        \includegraphics[width=\linewidth]{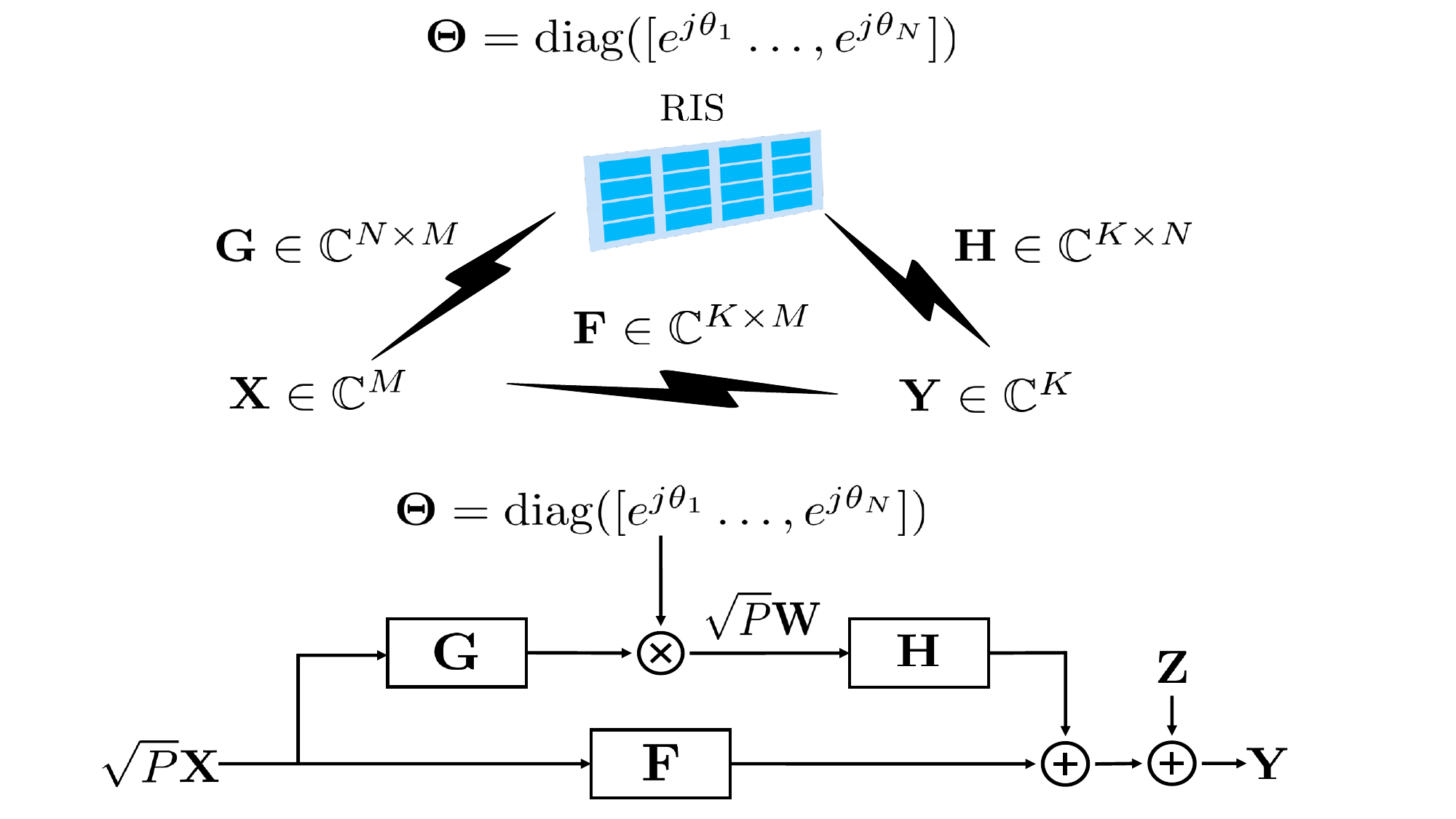}
        \caption{Channel model for a RIS assisted point-to-point MIMO communication system in which $(\mathbf{X},\mathbf{\Theta})$ jointly transmit as an input to the channel}
        \label{RIS}
\end{figure}

The RIS-aided communication scenario is as shown in Fig.~\ref{RIS}, where an RIS with $N$ reflective elements is deployed
between an $M$-antenna transmitter and a $K$-antenna receiver.
Let ${\mathbf{H}\in \mathbb{C}^{K \times N}}$, ${\mathbf{G}\in \mathbb{C}^{N \times M}}$, $\mathbf{F}\in \mathbb{C}^{K \times M}$ be the channel response matrices from the transmitter to the
RIS, from the RIS to the receiver, and from the transmitter to the receiver respectively, which are assumed to be fixed and known everywhere, i.e., at
the transmitter, the RIS controller, and the receiver. The channels can be estimated, e.g., using methods in \cite{chen2019channel, explicit_estimation}.

We assume that each element of the RIS combines all the transmitted signals at a single point and reflects it from this point to the receiver with an adjustable phase shift. The discrete-time channel model for the communication scenario is given by
\begin{equation}\label{general_model}
\mathbf{Y} = \sqrt{P}\left(\mathbf{H} {\boldsymbol {\Theta}} \mathbf{G}+\mathbf{F}\right) \mathbf{X} + \mathbf{Z},
\end{equation}
where $\mathbf{Y} \in \mathbb{C}^{K}$ is the received signal vector,
$\mathbf{X}\in \mathbb{C}^{M}$ is the transmit signal vector normalized by the power constraint $P$ so that
\begin{equation}
	\label{eq:x_constraint}
\mathbb{E}[\|\mathbf{X}\|_2^2]\leq 1,
\end{equation}
and $\mathbf{Z} \sim \mathcal{CN}(0, \sigma^2 \mathbf{I}_K)$ is the additive Gaussian noise vector at the receiver.
Here, ${\bm \Theta}$ is the $N\times N$ reflecting coefficient matrix induced by RIS taking the form of
\begin{equation}
	\label{eq:theta_constraint}
{\bm \Theta}={\text
{diag}}([e^{j\theta_1},e^{j\theta_2},\cdots,e^{j\theta_{N}}]),
\end{equation}
where ${ \theta}_m$ is the
phase-shifting coefficient of the $m$-th reflective element on the RIS and
${\text{diag}}(\cdot)$ denotes a diagonal matrix whose diagonal elements are
given by the corresponding entries of the argument.
Implicit in this discrete-time channel model is the assumption that the
transmitter and the RIS operate at the same symbol rate. The case in which the RIS operates at a slower rate is treated in Section \ref{sec:different_rate}.

An important special case of model \eqref{general_model} is when the direct path from the transmitter to the receiver is blocked, i.e. $\mathbf{F}=\mathbf{0}$. In this case, model \eqref{general_model} reduces to
\begin{equation}
\mathbf{Y} = \sqrt{P}\mathbf{H} \underbrace{{\boldsymbol {\Theta}} \mathbf{G} \mathbf{X}}_{\mathbf{W}}  + \mathbf{Z}. \label{systemmodel}
\end{equation}
Model \eqref{systemmodel} is easier to analyze. This paper begins by focusing on model \eqref{systemmodel}; the more general model \eqref{general_model} is treated in the second half of the paper. For convenience, we define $\mathbf{W} = \mathbf{\Theta G X}$ as shown in (\ref{systemmodel}) so that the output of the RIS is $\sqrt{P} \mathbf{W}$.

\subsection{RIS as Passive Beamformer}

In the conventional model, the RIS is used
as a passive beamformer to improve the overall transmission quality. In this case, the system design problem is simply that of finding $\mathbf{\Theta}$ as a function of $\mathbf{H}$ and $\mathbf{G}$ in order to
maximize the channel capacity from $\mathbf{X}$ to $\mathbf{Y}$. Mathematically, this maximization operation can be written as
\begin{equation}
\max_{\boldsymbol {\Theta}} \max_{p(\mathbf{x})} I\left(\mathbf{X}; \mathbf{Y}|\boldsymbol {\Theta}\right),
\end{equation}
subject to the constraints (\ref{eq:x_constraint}) and (\ref{eq:theta_constraint}).
In this case, the reflecting coefficients do not carry any information. The main function of the RIS is to enhance the beamforming gain from $\mathbf{X}$ to $\mathbf{Y}$, i.e. to increase the received SNR at the receiver.

Instead of using the RIS as a conventional beamformer, this paper studies the use of RIS for encoding information in the reflective coefficients themselves. Toward this end, we consider two separate scenarios.

\subsection{Joint Transmission Model}

The first scenario considered in this paper is the case in which
the RIS cooperates with the transmitter to transmit data.
This corresponds to the scenario in which the transmitter controls the RIS via a backhaul link,
and the overall channel is equivalent to a point-to-point channel from $\mathbf{(X, \Theta)}$ to $\mathbf{Y}$.

The capacity of such a channel, as a function of the transmitted power $P$, can be written as
\begin{equation}
C(P) = \max_{p(\mathbf{x}, \boldsymbol{\Theta} )} I(\mathbf{X},\boldsymbol{\Theta}; \mathbf{Y}),
\end{equation}
subject to the constraints (\ref{eq:x_constraint}) and (\ref{eq:theta_constraint}).
This is referred to as the \emph{joint transmission} model and is considered in the first part of this paper.

Computing $I(\mathbf{X},\boldsymbol {\Theta}; \mathbf{Y})$ is not an easy task as it involves a multi-dimensional integration over the probability density functions (pdf) of $\mathbf{X}$ and $\boldsymbol {\Theta}$, which can only be done numerically.
To obtain some insight into the problem, in this paper, we characterize the DoF of this channel model, which is defined as the pre-log factor of the rate as $P\rightarrow \infty$, i.e.
\begin{equation}
\mathrm{DoF}_{(\mathbf{X, \Theta})} \triangleq \lim_{P \rightarrow \infty} \frac{C(P)}{\log P}.
\end{equation}

\subsection{Multiple Access Model}

The second scenario considered in this paper is the case in which the
transmitter and the RIS send independent data to the receiver without
cooperation between the two. For example, this can correspond
to a scenario in which the RIS is part of a sensor network, which needs to send the
environmental information collected by the sensors to the receiver.
This channel model is referred to as the \emph{multiple access} model.

Although the channel
remains the same as in the joint transmission case, because
there is no cooperation between $\mathbf{X}$ and $\boldsymbol{\Theta}$ in
the multiple access model,
the input distributions are constrained to be the product of the marginal
distributions $p(\mathbf{X})$ and $p(\mathbf{\Theta})$. 
Denote the achievable rate of $\mathbf{X}$ from the transmitter as $R_{\mathbf{X}}$ and
the achievable rate of $\boldsymbol{\Theta}$ from the RIS as
$R_{\boldsymbol{\Theta}}$. The capacity region of the multiple access channel is
the closure of the convex hull of all ($R_{\mathbf{X}}$, $R_{\boldsymbol{\Theta}}$)
satisfying
\begin{eqnarray}
	R_{\mathbf{X}} & \leq & I(\mathbf{X}; \mathbf{Y}|\boldsymbol{\Theta}), \\
	R_{\boldsymbol{\Theta}} & \leq & I(\boldsymbol{\Theta}; \mathbf{Y}|\mathbf{X}), \\
	R_{\mathbf{X}}+R_{\boldsymbol{\Theta}}  & \leq & I(\mathbf{X},\boldsymbol{\Theta}; \mathbf{Y}).
\end{eqnarray}
for some joint distributions $p(\mathbf{X})p( \boldsymbol{\Theta})$ 
subject to the constraints (\ref{eq:x_constraint}) and (\ref{eq:theta_constraint}).

The capacity region is a function of the transmit power $P$.
A DoF pair $(\mathrm{DoF}_\mathbf{X}, \mathrm{DoF}_{\boldsymbol{\Theta}})$ is
achievable if there exists a sequence of rate pairs
$(R_{\mathbf{X}}(P), R_{\boldsymbol{\Theta}}(P))$
in the capacity region, as a function of $P$, such that
\begin{eqnarray}
\mathrm{DoF}_\mathbf{X} & = & \displaystyle \lim_{P \rightarrow \infty} \frac{R_{\mathbf{X}}(P)}{\log P}, \\
\mathrm{DoF}_{\boldsymbol{\Theta}} & = & \displaystyle \lim_{P \rightarrow \infty} \frac{R_{\boldsymbol{\Theta}}(P)}{\log P}.
\end{eqnarray}
The DoF region $\{(\mathrm{DoF}_\mathbf{X}, \mathrm{DoF}_{\boldsymbol{\Theta}})\}$
of the multiple access model
is defined to be the closure of the convex hull of all achievable DoF pairs.

The goal of this paper is to characterize the DoF of the joint transmission and the DoF region of the multiple access model.

\section{DoF of RIS System without Direct Path}
\label{sec:no_direct_path}

We begin by considering the RIS model \eqref{systemmodel} without the direct
path. The aim is to characterize both the DoF of the joint transmission case,
where
$\mathbf{X}$ and $\mathbf{\Theta}$ jointly transmit information to the receiver
$\mathbf{Y}$, and the multiple access case, where $\mathbf{X}$ and
$\mathbf{\Theta}$ are independent users.

A key observation is that the RIS
channel model is closely related to the channel with phase noise. To illustrate this idea, we start with the single-input single-output (SISO) additive white Gaussian noise (AWGN) channel as an example, then move onto the MIMO case.

\subsection{SISO Channel}
\label{sec:siso}
Consider the special case of $M=N=K=1$, for which the signal model \eqref{systemmodel} reduces to
\begin{eqnarray}
\label{model_siso}
Y & = & \sqrt{P} e^{j\theta} X + Z \\
  & = & \sqrt{P} |X| e^{j(\angle X +\theta)} + Z,
\end{eqnarray}
where $Z \sim \mathcal{CN}(0,\sigma^2)$, and the channel gain is omitted for simplicity.
The input constraints are $\theta \in (-\pi,\pi]$ and $\mathbb{E}|X|^2 \le 1$.
Here, we use $|\cdot|$ to denote the amplitude and $\angle$ as the phase of a complex number.

\begin{figure}[t]
\centering
\includegraphics[width=0.97\linewidth]{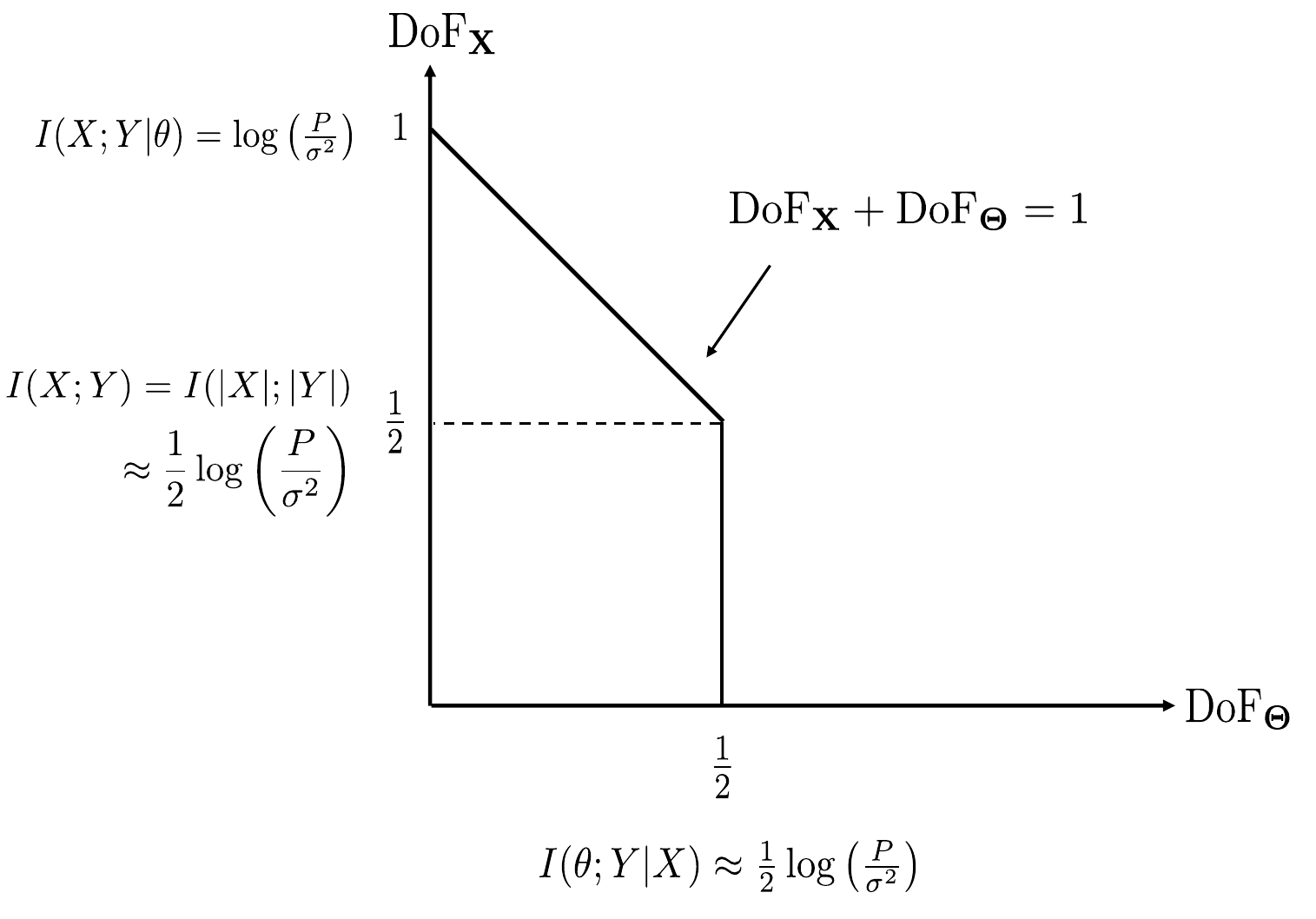}
\caption{DoF region of the SISO multiple access channel model}\label{fig:region1}
\end{figure}

The capacity for joint transmission is the maximum mutual information between the input $(X,\theta)$ and the output $Y$. Observe that any phase of $Y$ that can be synthesized by $X$ and $\theta$ can be equivalently produced by $X$ itself with no help from $\theta$.
Thus, the capacity of joint transmission is exactly the same as the conventional AWGN channel.

To gain additional insight, we can also compute this capacity in an alternative way as below. 

From the chain rule, the mutual information $I(X,\theta;Y)$ can be decomposed as
\begin{equation}\label{polar}
I(X,\theta;Y) 
			   = I(X;Y) + I(\theta;Y | X).
\end{equation}
Assume that we choose $p(\theta) \sim \operatorname{Unif}(-\pi, \pi]$
and let $X$ and $\theta$ be independent.
Consider the first term $I(X;Y)$.
Since $\theta$ is completely unknown,
the channel from $X$ and $Y$ is in effect a non-coherent channel
\cite{katz2004capacity}, in which $\theta$ can be thought of as phase noise.
For such a phase-noise channel, information can only be conveyed through the magnitude,
so we have
\begin{equation}
I(X;Y) = I(|X|;|Y|).
\end{equation}
The optimal distribution of the magnitude $|X|$ for maximizing $I(|X|; |Y|)$ is
known to be discrete \cite[Theorem 5]{katz2004capacity}. However, setting $X$
as Gaussian can be shown to give the same DoF
\cite{blachman1953comparison}. The achievable rate using Gaussian input in the
high SNR regime can be approximated as (\cite{goebel2011calculation,katz2004capacity}):
\begin{equation} \label{eq:region2}
I(|X|;|Y|) \approx \frac{1}{2}\log\left(\frac{P}{\sigma^2}\right)-0.69.
\end{equation}
Thus, the $I(X;Y)$ term achieves a DoF of $1/2$. 

Now, consider the second term $I(\theta; Y|X)$, where $X$ is known and can be
considered as part of the channel gain.
In this case, only the phase $\theta$ carries information. This is known as
the constant-intensity channel, or constant-envelope channel, or ring
modulation \cite{blachman1953comparison,phase1993}.
Intuitively, the input is a phase-shift keying (PSK) modulation.
The asymptotic approximation of the capacity of such a channel is found in
the high SNR limit in \cite[eqn. (25)]{goebel2011calculation} and is given by
\begin{align}\label{eq0}
\max_{p(\theta)} I(\theta;Y|X)
& \approx \frac{1}{2}\log\left(\frac{P |X|^2}{\sigma^2}\right)+1.1 \nonumber \\
& = \frac{1}{2}\log\left(\frac{P}{\sigma^2}\right)+ \log|X| + 1.1,
\end{align}
where the optimal distribution of \(\theta\) is uniform  in $(-\pi,\pi]$.
Since we have $\mathbb{E}|X|^2 \le 1$, the $\log|X|$ term cannot scale with SNR. Thus, the $I(\theta; Y|X)$ term also achieves a DoF of $1/2$.

Combining the two terms, we see that for the SISO channel, the joint
transmission of $(X, \theta)$ achieves the same DoF as that of a
conventional AWGN channel without RIS.

For the AWGN channel without RIS, the DoF of 1 can be thought of as being comprised of $1/2$ DoF in magnitude of  $X$ and $1/2$ DoF in phase of $\theta$. But with RIS as part of the channel, there is a $1/2$ DoF overlap between
the phase of the input $X$ and the phase of the RIS $\theta$. Thus, the joint transmission
of $X$ and $\theta$ achieves the same DoF with contributions of $1/2$ DoF in magnitude of $X$ and $1/2$ DoF in phase of either $X$ or $\theta$.

The above argument is useful because it can be readily applied to the multiple access setting, in which $X$ and $\theta$ have to be independent. The resulting derivation gives the DoF region of the SISO multiple access channel 
as shown in Fig.~\ref{fig:region1}.

\subsection{Joint Transmission in MIMO Channel with $(\mathbf{X},\mathbf{\Theta})$ as Input}
\label{sec:joint}

We now consider the MIMO case for the joint transmission of $\mathbf{X}$ and
$\mathbf{\Theta}$, and with arbitrary $M$, $N$, and $K$, and compute the DoF of
$I(\mathbf{X},\boldsymbol{\Theta};\mathbf{Y})$, but assuming that there is no
direct path between $\mathbf{X}$ and $\mathbf{Y}$.
This result will be generalized to the case with
direct path later, but the model without the direct path is also of
independent interest, because in many cases the RIS is deployed
where the direct path between the transmitter and the receiver is blocked.

The main result below, which is for the point-to-point case with $(\mathbf{X},
\mathbf{\Theta})$ jointly transmitting as the input to channel, is also used later
as a tool for showing the DoF region of the multiple access case.

\begin{theorem}\label{theoremNneqK}
Consider the channel model
\begin{equation}
\mathbf{Y}=\sqrt{P}\mathbf{H W}  + \mathbf{Z} = \sqrt{P}{\mathbf{H} \boldsymbol {\Theta} \mathbf{G}} \mathbf{X}  + \mathbf{Z},
\end{equation}
with $(\mathbf{X,\Theta})$ as the input where $\mathbf{X}\in \mathbb{C}^M$ has a power constraint $\mathbb{E}[\|\mathbf{X}\|_2^2]\leq 1$ and ${\boldsymbol{{\Theta}}}={\text{diag}}([e^{j\theta_1},e^{j\theta_2},\cdots,e^{j\theta_{{N}}}])$, and $\mathbf{Y}\in\mathbb{C}^K$  as the output, and  $\mathbf{Z} \sim\mathcal{CN}(0,\mathbf{I}_{K})$ as the noise.
For almost all matrices $\mathbf{G} \in \mathbb{C}^{N\times M}$, $\mathbf{H} \in \mathbb{C}^{K\times N}$, the DoF of the channel is
\begin{equation}
\mathrm{DoF}_{(\mathbf{X}, \mathbf{\Theta})} = \min \left(M+\frac{N}{2}-\frac{1}{2}, N, K \right).
\label{eq:HW_dimension}
\end{equation}
This DoF can be achieved using independent $\mathbf{X}$ and $\mathbf{\Theta}$.
When $K\geq \min\left(M+\frac{N}{2}-\frac{1}{2},N\right)$, any of the following input distributions with independent $\mathbf{X}$ and $\mathbf{\Theta}$ would achieve the DoF in \eqref{eq:HW_dimension}.
\begin{enumerate}
\item $\mathbf{X} \sim \mathcal{CN}(0, \mathbf{I}_{M})$ and $\boldsymbol{\Theta}$
	 with independently and identically distributed (i.i.d.) 
	$\theta_i$'s and $\theta_i \sim \operatorname{Unif}(-\pi, \pi]$.
\item The distribution of $\mathbf{X}$ is chosen such that the first $M-1$ elements are i.i.d. complex Gaussian, i.e., $\mathbf{X}_{M-1}\sim \mathcal{CN}(0,\mathbf{I}_{M-1})$,
and $X_{M}$ is chosen to be real with a chi-squared distribution with 2 degrees of freedom.
The distribution of $\boldsymbol{\Theta}$ is such that $\theta_i$'s are i.i.d. with $\theta_i \sim \operatorname{Unif}(-\pi, \pi]$.
\item The distribution of $\mathbf{X}$ is i.i.d. complex Gaussian for all $M$ elements, i.e., $\mathbf{X}\sim \mathcal{CN}(0,\mathbf{I}_{M})$. For $\mathbf{\Theta}$,
the first ${N}-1$ elements of $\boldsymbol{\Theta}$ are chosen as i.i.d. with $\theta_i \sim \operatorname{Unif}(-\pi, \pi], ~i=1,\ldots, N-1$, and the last element of $\boldsymbol{\Theta}$ is fixed as 1, i.e., $\theta_N=1$.
\end{enumerate}
When $K<\min\left(M+\frac{N}{2}-\frac{1}{2},N\right)$, the DoF is similarly achieved with the above distributions modified by setting some of the elements in $\mathbf{X}$ and $\mathbf{\Theta}$ to be deterministic. 
\end{theorem}

\begin{IEEEproof}
The proof is
based on an equivalence between information dimension and the DoF of an additive noise channel. Toward this end, we first compute the information dimension of the output $\mathbf{W}$ at the RIS. This is accomplished by building a connection between the channel model with RIS and the MIMO channel with phase noise. We then compute the information dimension of $\mathbf{HW}$ by studying the behavior of information dimension under projection.
The details of the proof are in Appendix \ref{app:proof_DoFp2p}.
\end{IEEEproof}

The statements of this and also subsequent theorems refer to
``almost all'' matrices, by which we mean all matrices except ones lying on
some lower dimensional algebraic hypersurfaces (e.g., the low-rank matrices).
If the entries of the matrices are drawn i.i.d.\ according to some absolutely continuous
distribution, then the statement of the theorem holds almost surely.

We now give an interpretation of the DoF result (\ref{eq:HW_dimension}) in
Theorem \ref{theoremNneqK}.

The channel model has two inputs: $\mathbf{X}$ of
dimension $M$, and $\mathbf{\Theta}$ of dimension $N$. Because
$\mathbf{\Theta}$ has phase control only, its information dimension is
$\frac{N}{2}$. Further, $\mathbf{X}$ and $\mathbf{\Theta}$ have $1/2$
dimension overlap, due to the product form of the channel model, which leaves
a common phase ambiguity between $\mathbf{X}$ and $\mathbf{\Theta}$.
In other words, it is not possible to resolve the phases of $\mathbf{X}$ and
$\mathbf{\Theta}$ individually, but only the phase of their product.
For this reason, the total input dimension is
$M+\frac{N}{2}- \frac{1}{2}$.

Moreover, the channel has output $\mathbf{Y}$ of
dimension $K$, and the RIS acts as a relay with a bottleneck $\mathbf{W}$ of
dimension $N$. The overall DoF must be the minimum of the input dimension, the
relay dimension, and the output dimension. This is why the multiplexing gain is
$\min (M+\frac{N}{2}-\frac{1}{2}, N, K)$.

The key observation here is the $1/2$ dimension overlap between $\mathbf{X}$
and $\mathbf{\Theta}$. It is for this reason that the overall DoF
can be achieved with not only a choice of input distributions with
$\mathbf{X} \sim \mathcal{CN}(0, \mathbf{I}_{ M})$ and $\boldsymbol{\Theta}$
with $\theta_i \sim \operatorname{Unif}(-\pi, \pi]$, but also with $1/2$
dimension removed from either $\mathbf{X}$ or $\mathbf{\Theta}$.
In particular, Theorem \ref{theoremNneqK} shows that fixing
the phase of the last element of either $\mathbf{X}$ or $\boldsymbol{\Theta}$
still allows the same DoF to be achieved.
In fact, due to symmetry, the same result holds true if we fix any of one phase value in either $\mathbf{X}$ or $\mathbf{\Theta}$.

An important implication of Theorem \ref{theoremNneqK} is that
as compared to a MIMO channel with $M$ antennas at the input and $K$ antennas at the output achieving a DoF of $\min(M,K)$, an RIS can improve the overall DoF if $K > M$, and $N > \min(M,K)$.
In essence, the phase shifts at the RIS can act as part of the input to effectively increase the total input dimensions. Thus, by making the input data available to the RIS, we are enabling the RIS to help the transmitter modulate the data stream via the phase shifts. Assuming that $N$ is much larger than $M$ and $K$ as in typical deployment scenarios and $K>M$, the maximum additional number of data streams that the RIS can provide is
$\min\left(K-M, \frac{N}{2}-\frac{1}{2}\right)$.
Note that if the RIS is used merely as a beamformer or as a reflector, it cannot improve the DoF. Thus, modulating data through phase shifts has significant benefits.

A practical case is when the transmitter has a single radio-frequency (RF) chain. This is akin to using the RIS to emulate a MIMO transmitter, i.e., a single-antenna active transmitter together with an RIS can be jointly configured to act as a MIMO array. Such a system can be considered as a low-cost alternative to a fully digital MIMO system. In this case, we have $M=1$ and the signal model is given by
\begin{equation}
\mathbf{Y} = \mathbf{H} {\boldsymbol {\Theta}} \mathbf{g} X  + \mathbf{W}, \label{systemmodel5}
\end{equation}
Assuming an RIS with large $N$, Theorem \ref{theoremNneqK} implies that the overall DoF is $K$. One possible way of achieving this DoF is to use the amplitude of $X$ to modulate $\frac{1}{2}$ DoF and to use a RIS of size at least $N=2K-1$ to modulate the remaining $(K-\frac{1}{2})$ DoF.



We remark that the DoF of the RIS system is quite different from that of a relay channel.
For the relay channel model with $N$ active antennas at the relay and without a direct path, if the relay
is also able to encode information, the DoF of the overall channel is simply $\min(N,K)$,
assuming no unit modulus constraint on the relay output.
In contrast, the DoF of the RIS system, i.e., (\ref{eq:HW_dimension}), is achieved with passive RIS elements
with adjustable phases. 

\subsection{MIMO Channel with Deterministic $\mathbf{X}$}
\label{sec:knownX}

We can carry the example in the previous section one step further and consider
a system in which the information is modulated only from the RIS via $\mathbf{\Theta}$ and not by the
transmitter at all.
In this case, the transmitter sends a deterministic signal $\mathbf{X}$ (which carries no
information) and the channel model becomes
\begin{align}\label{knownX}
\mathbf{Y}&=\sqrt{P}\mathbf{H}\boldsymbol{\Theta}\mathbf{G}\mathbf{X}+ \mathbf{Z} \\
&=\sqrt{P}\mathbf{H}\mathrm{diag}(\mathbf{G}\mathbf{X})\boldsymbol{\Phi}+ \mathbf{Z} \\
&=\sqrt{P}\bar{\mathbf{H}} \boldsymbol{\Phi}+ \mathbf{Z},
\end{align}
where $\boldsymbol{\Phi}=[e^{j\theta_1},\ldots,e^{j\theta_{N}}]^{T}$
is the vector form of the matrix $\boldsymbol{\Theta}$ and
is now the information-carrying input signal from the RIS, and
$\bar{\mathbf{H}}=\mathbf{H}\mathrm{diag}(\mathbf{G}\mathbf{X})$ is
the equivalent channel.
If we assume that $\mathbf{X}$ is such that none of the diagonal elements of
$\mathrm{diag}(\mathbf{G}\mathbf{X})$ are zero, then
\eqref{knownX} is equivalent to a MIMO channel with constant-amplitude
and continuous-phase inputs on each of its transmit antennas.


The above channel model can also be thought of as a special case of
Theorem \ref{theoremNneqK} but with the input distribution restricted to
having a deterministic $\mathbf{X}$. The capacity of this channel is
\begin{align}
\max_{p(\boldsymbol{\Theta} )} I(\mathbf{X},\mathbf{\Theta}; \mathbf{Y})
= \max_{p(\boldsymbol{\Phi} )} I(\mathbf{\Phi}; \mathbf{Y}).
\end{align}
Unlike in Theorem \ref{theoremNneqK} though, because $\mathbf{X}$ is
deterministic, there is no $1/2$ DoF overlap between $\mathbf{X}$ and $\mathbf{\Theta}$,
and the DoF of the resulting channel is as follows:


\begin{theorem}\label{theorem_knownX}
Consider the channel model
\begin{equation}
\mathbf{Y}= \sqrt{P}\bar{\mathbf{H}} \boldsymbol{\Phi}+ \mathbf{Z},
\end{equation}
with $\boldsymbol{\Phi}=[e^{j\theta_1},\ldots,e^{j\theta_{N}}]^{T}$ as the input,
$\bar{\mathbf{H}}=\mathbf{H}\mathrm{diag}(\mathbf{G}\mathbf{X})$ as the channel,
$\mathbf{Y}\in\mathbb{C}^K$ as the output, and $\mathbf{Z} \sim\mathcal{CN}(0,\mathbf{I}_{K})$ as the noise.
For almost all matrices $\mathbf{H} \in \mathbb{C}^{K\times N}$,
$\mathbf{G} \in \mathbb{C}^{N\times M}$, and
for a deterministic $\mathbf{X}$ such that $\mathrm{diag}(\mathbf{G}\mathbf{X})$ has nonzero elements,
the DoF of the channel is
\begin{equation}
\mathrm{DoF}_{\mathbf{\Theta}} = \min \left(\frac{N}{2}, K \right).
\label{eq:knownX_dimension}
\end{equation}
\end{theorem}

\begin{IEEEproof}
Please see Appendix \ref{app:proof_DoFknownx}.
\end{IEEEproof}

Intuitively, modulating information through the phases of an RIS with $N$
passive elements alone and deterministic $\mathbf{X}$ can already achieve a DoF
of up to $N/2$. Comparing this result with the case of modulating information
through both the RIS and a single-antenna transmitter with $M=1$, the latter
case achieves a DoF of 
$\min\left(\frac{N+1}{2},K\right)$. The difference of $1/2$ DoF is due to the lack
of amplitude control when $\mathbf{X}$ is deterministic.


\subsection{Multiple Access Channel with $\mathbf{X}$ and $\mathbf{\Theta}$ as Inputs}
\label{sec:mac}

Thus far, we have considered the cases in which the transmitter and the RIS
cooperate in sending a single message to the receiver. In this section, we
consider the scenario in which the transmitter and the RIS
send independent messages to the receiver.
Recall that the main difference between the two cases is that the input
distribution for the multiple access channel must be a product of the
marginals.  Fortunately, the distribution that achieves the DoF in
Theorem \ref{theoremNneqK} is already in this form.  This gives the following
characterization of the DoF region of the multiple access channel.

\begin{figure*}
        \centering
        \subfigure[
	$\frac{N}{2} \geq M-\frac{1}{2}$ and $K\geq M+\frac{N}{2}-\frac{1}{2}$]{\includegraphics[height=0.21\textheight,keepaspectratio]{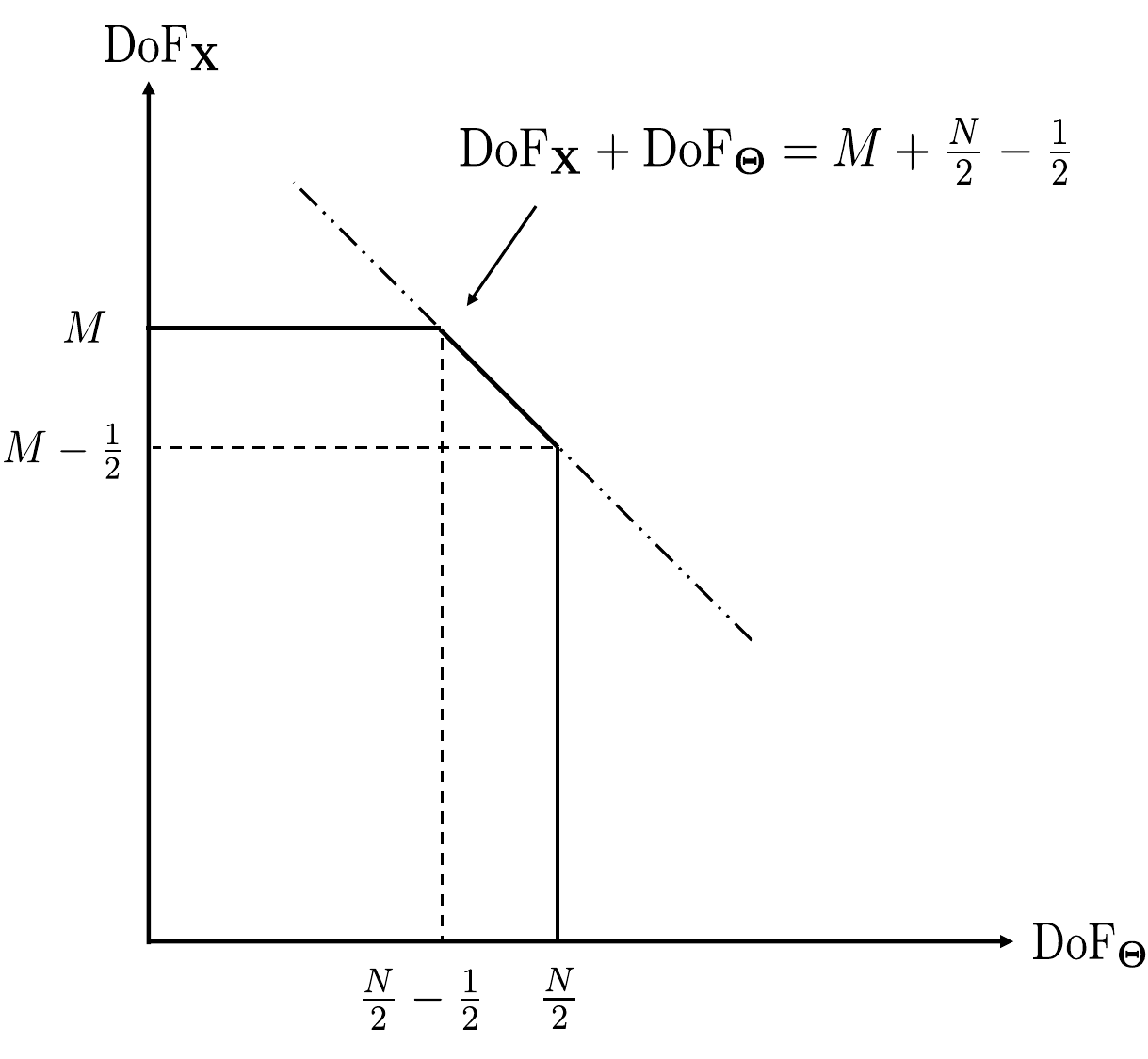}}
        \hspace{2.5mm}
	\subfigure[$M \leq K\leq \frac{N}{2}$ and $K\leq M+\frac{N}{2}-\frac{1}{2}$]{\includegraphics[height=0.21\textheight,keepaspectratio]{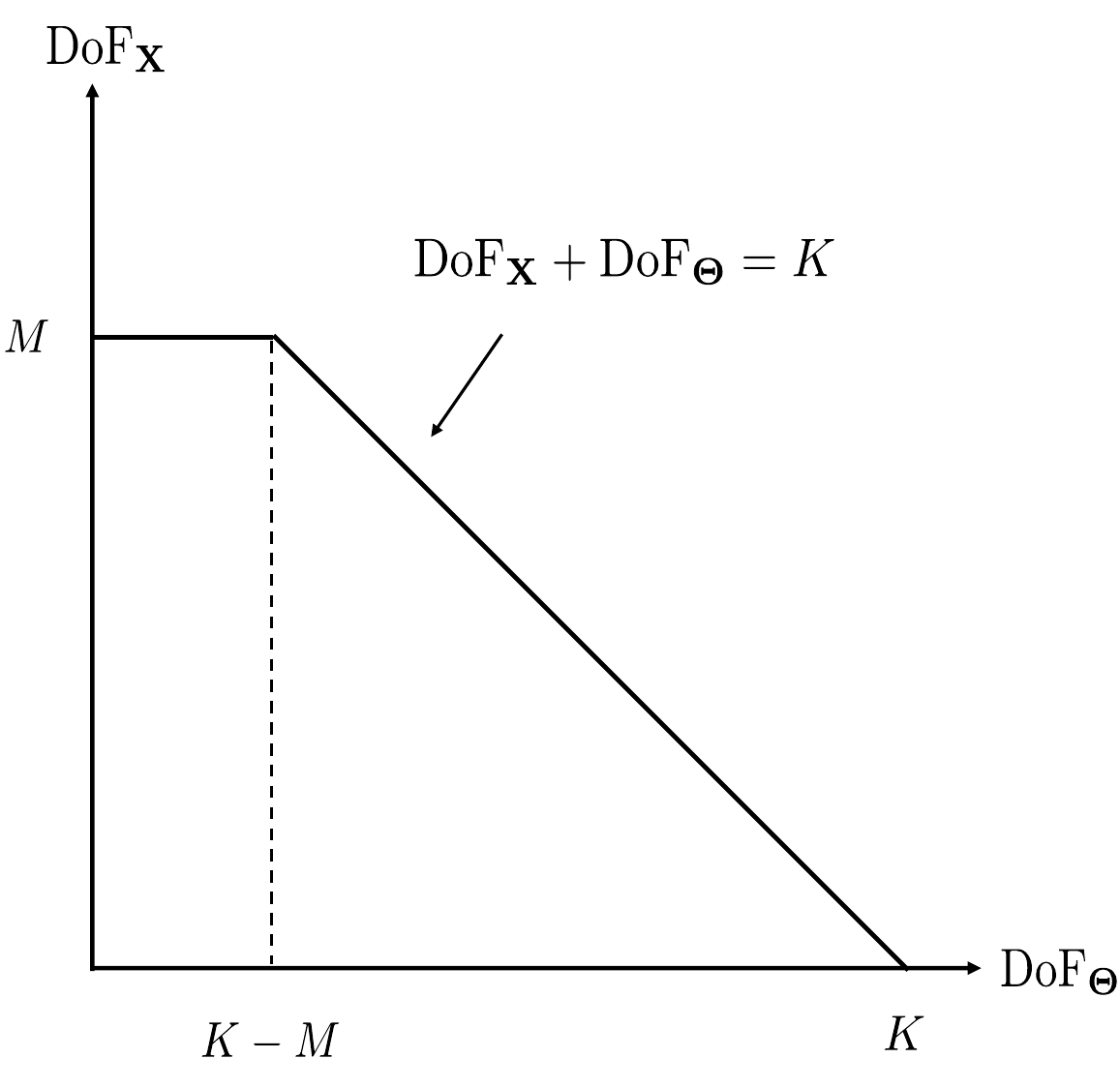}}
        \hspace{2.5mm}
	\subfigure[$\frac{N}{2}+\frac{1}{2} \le M \le N \le K$]{\includegraphics[height=0.21\textheight,keepaspectratio]{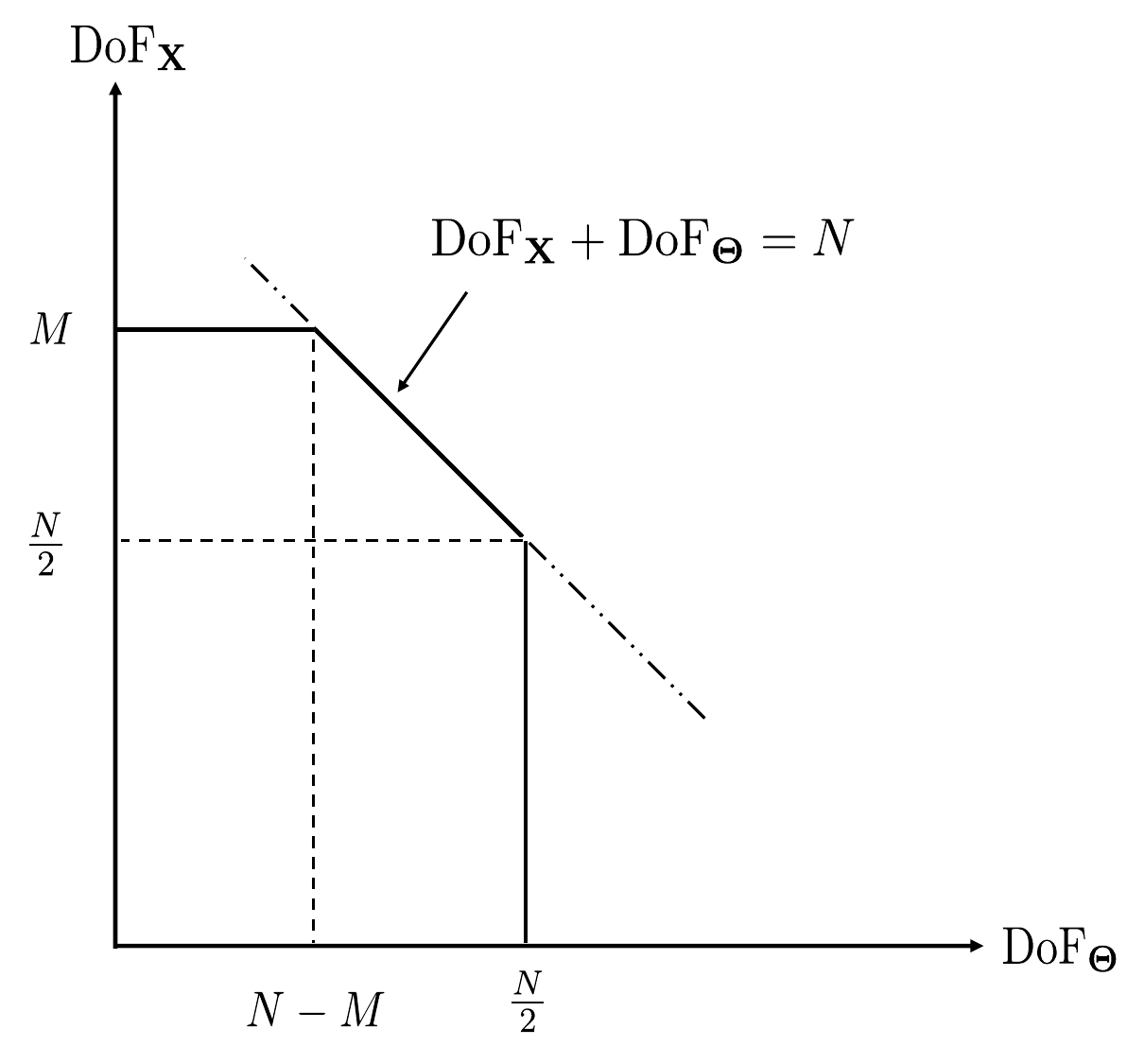}}
        \caption{Examples of DoF region for the multiple access channel without a direct path under different values of $M$, $N$ and $K$.}
        \label{region2}
\end{figure*}

\begin{theorem}\label{DoF_region}
Consider the multiple access channel model
\begin{align}
\mathbf{Y}=\sqrt{P}\mathbf{H}\boldsymbol{\Theta}\mathbf{G}\mathbf{X}+ \mathbf{Z},
\end{align}
with $\mathbf{X}$ and $\boldsymbol{\Theta}$ as the two independent inputs,
$\mathbf{Y}\in\mathbb{C}^K$ as the output, and $\mathbf{Z} \sim\mathcal{CN}(0,\mathbf{I}_{K})$ as the noise.
Here, $\mathbf{X}\in\mathbb{C}^M$ has a power constraint $\mathbb{E}[\|\mathbf{X}\|_2^2]\leq 1$, and ${\boldsymbol{{\Theta}}}={\text{diag}}([e^{j\theta_1},e^{j\theta_2},\cdots,e^{j\theta_{{N}}}])$ has a unit modulus constraint.
For almost all matrices $\mathbf{G} \in \mathbb{C}^{N\times M}$ and
$\mathbf{H} \in \mathbb{C}^{K\times N}$, the DoF region of the multiple access
channel is given by the set of DoF pairs $(\mathrm{DoF}_{\mathbf{X}}, \mathrm{DoF}_{\boldsymbol{\Theta}})$ that satisfy
\begin{eqnarray}
	\mathrm{DoF}_{\mathbf{X}} & \leq & \min(M, N, K), \label{eq:DoF_region_1} \\
	\mathrm{DoF}_{\boldsymbol{\Theta}} & \leq & \min\left(\frac{N}{2},K\right), \label{eq:DoF_region_2} \\
	\mathrm{DoF}_{\mathbf{X}}+\mathrm{DoF}_{\boldsymbol{\Theta}}  & \leq & \min \left(M+ \frac{N}{2}-\frac{1}{2}, N, K \right).  \label{eq:DoF_region_3}
\end{eqnarray}
\end{theorem}
\begin{IEEEproof}
First, we show the converse.
To bound $I(\mathbf{X};\mathbf{Y}|\mathbf{\Theta})$, observe that conditioned on $\mathbf{\Theta}$, the channel from $\mathbf{X}$ to $\mathbf{Y}$ is equivalent to a MIMO channel with a channel matrix $\mathbf{H}_{\mathrm{eq}}=\mathbf{H}\boldsymbol{\Theta}\mathbf{G}$.
Thus, $\mathrm{DoF}_{\mathbf{X}}$ is upper bounded by the rank of the equivalent channel as follows: 
\begin{equation}
	\mathrm{DoF}_{\mathbf{X}} \leq \mathrm{rank} (\mathbf{H}_{\mathrm{eq}}) =\min (M,N,K).
\end{equation}

Similarly, $\mathrm{DoF}_{\boldsymbol{\Theta}}$ is upper bounded by the input dimension $\frac{N}{2}$ and the output dimension $K$, i.e.,
\begin{equation}
\mathrm{DoF}_{\boldsymbol{\Theta}} \leq \min\left(\frac{N}{2},K\right).
\end{equation}

Finally, the sum DoF is upper bounded
by allowing cooperation between $\mathbf{X}$ and $\boldsymbol{\Theta}$. By Theorem \ref{theoremNneqK}, we have
\begin{equation}
\mathrm{DoF}_{\mathbf{X}}+\mathrm{DoF}_{\boldsymbol{\Theta}} \leq \min \left(M+ \frac{N}{2}-\frac{1}{2}, N, K \right).
\end{equation}

Next, we show achievability. The key is to recognize that the sum DoF as characterized in Theorem \ref{theoremNneqK} is achieved with several choices of independent distributions $p(\mathbf{X})p(\mathbf{\Theta})$. Thus, this same sum DoF is also achievable in the multiple access setting. The different choices of the independent distributions can be used to achieve the different corner points of the DoF region outer bound above.



Choose the distribution of $\mathbf{X}$ and $\boldsymbol{\Theta}$ as in Case 2) of Theorem \ref{theoremNneqK}.
We decode $\mathbf{X}$ first, then $\mathbf{\Theta}$, which corresponds to decomposing the sum rate as
\begin{align}
I(\mathbf{X}, \boldsymbol {\Theta};\mathbf{Y})
=I\left(\boldsymbol {\Theta};\mathbf{Y}|\mathbf{X}\right) + I(\mathbf{X};\mathbf{Y}).
\end{align}
The first term $I\left(\boldsymbol {\Theta};\mathbf{Y}|\mathbf{X}\right)$ gives
rise to an achievable $\mathrm{DoF}_{\boldsymbol{\Theta}}$ as (\ref{eq:DoF_theta_1}) below according to Theorem \ref{theorem_knownX}. The achievable $\mathrm{DoF}_{\mathbf{X}}$ can be derived by subtracting  $ \mathrm{DoF}_{\boldsymbol{\Theta}} $ from the achievable sum DoF.
This gives the following achievable DoF pair:
\begin{align}
& \mathrm{DoF}_{\boldsymbol{\Theta}} = \min\left(\frac{N}{2},K\right), \label{eq:DoF_theta_1}\\
& \mathrm{DoF}_{\mathbf{X}} = 
 \min \left(M+ \frac{N}{2}-\frac{1}{2}, N, K \right)
 - \min\left(\frac{N}{2},K\right).
\end{align}

Next, choose the distributions as in Case 3) in Theorem \ref{theoremNneqK}, which achieve the same sum DoF of $\min \left(M+ \frac{N}{2}-\frac{1}{2}, N, K \right)$. 
We now decode $\mathbf{\Theta}$ first, then $\mathbf{X}$, which corresponds to decomposing the sum rate
$I(\mathbf{X}, \boldsymbol {\Theta};\mathbf{Y})$ as
\begin{align}
I(\mathbf{X}, \boldsymbol {\Theta};\mathbf{Y})
=I\left(\mathbf{X};\mathbf{Y}|\boldsymbol {\Theta}\right) + I(\boldsymbol {\Theta};\mathbf{Y}).
\end{align}
The first term $I\left(\mathbf{X};\mathbf{Y}|\boldsymbol {\Theta}\right)$ gives rise to an achievable $\mathrm{DoF}_{\mathbf{X}}$ as in (\ref{eq:DoF_X_2}) below, because the equivalent channel is just a point-to-point MIMO channel with rank $\min(M,N,K)$.
The achievable $\mathrm{DoF}_{\boldsymbol{\Theta}}$ can be derived by subtracting  $ \mathrm{DoF}_{\mathbf{X}} $ from the achievable sum DoF.
This gives the following achievable DoF pair:
\begin{align}
& \mathrm{DoF}_{\mathbf{X}} = \min(M, N, K), \label{eq:DoF_X_2}\\
& \mathrm{DoF}_{\boldsymbol{\Theta}} =
 \min \left(M+ \frac{N}{2}-\frac{1}{2}, N, K \right) -
 \min(M, N, K).
\end{align}

Finally by time-sharing, we get the achievability of the entire region (\ref{eq:DoF_region_1})-(\ref{eq:DoF_region_3}).
\end{IEEEproof}

Depending on the relative values of $M$, $N$ and $K$, the DoF region may take
different shapes as shown by several examples in Fig.~\ref{region2}.
Case (a) is an example where there is a large number of receive antennas (i.e., large $K$).
In this case, the DoF region is constrained by the transmitter dimension $M$
and the RIS dimension $N/2$, but because of the $1/2$ dimension overlap,
the DoF region has a pentagon shape.
Case (b) is an example where the DoF region is constrained by the receiver dimension $K$.
Case (c) is an example where the DoF region is constrained by the RIS dimension $N$.
In all three examples, the RIS can transmit additional data streams to the receiver
on top of the $M$ data streams from the transmitter. They show the benefit of
modulating information through the phases of the RIS.

As an additional remark, the input distributions that achieve the sum DoF in the proof of
Theorem \ref{DoF_region} are i.i.d.\ across all the elements of $\mathbf{X}$
and $\boldsymbol{\Theta}$, except the overlapping dimension. This implies that
this result can be generalized to the scenarios in which $\mathbf{X}$ or
$\boldsymbol{\Theta}$ can consist of multiple users.
This corresponds to a system with multiple transmitters or multiple RIS's, i.e., different
components of $\mathbf{X}$ and $\mathbf{\Theta}$ can represent different users or RISs,
respectively.



\section{DoF of RIS System with Direct Path}
\label{sec:direct_path}

We now generalize the analysis of the previous section to the channel model where a direct path exists 
between the transmitter and the receiver. 
Without the direct path, the previous analysis shows that there is a global phase ambiguity resulting in a loss of $1/2$ DoF in both the point-to-point and the multiple access case. In this section, we show that this loss of DoF due to phase ambiguity can be recovered when a direct path is present.


\subsection{Joint Transmission with $(\mathbf{X},\mathbf{\Theta})$ as Input}

\begin{figure*}
        \centering
	\subfigure[$ \frac{N}{2}\geq M$ and $K\geq M+\frac{N}{2}$]{\includegraphics[height=0.2\textheight,keepaspectratio]{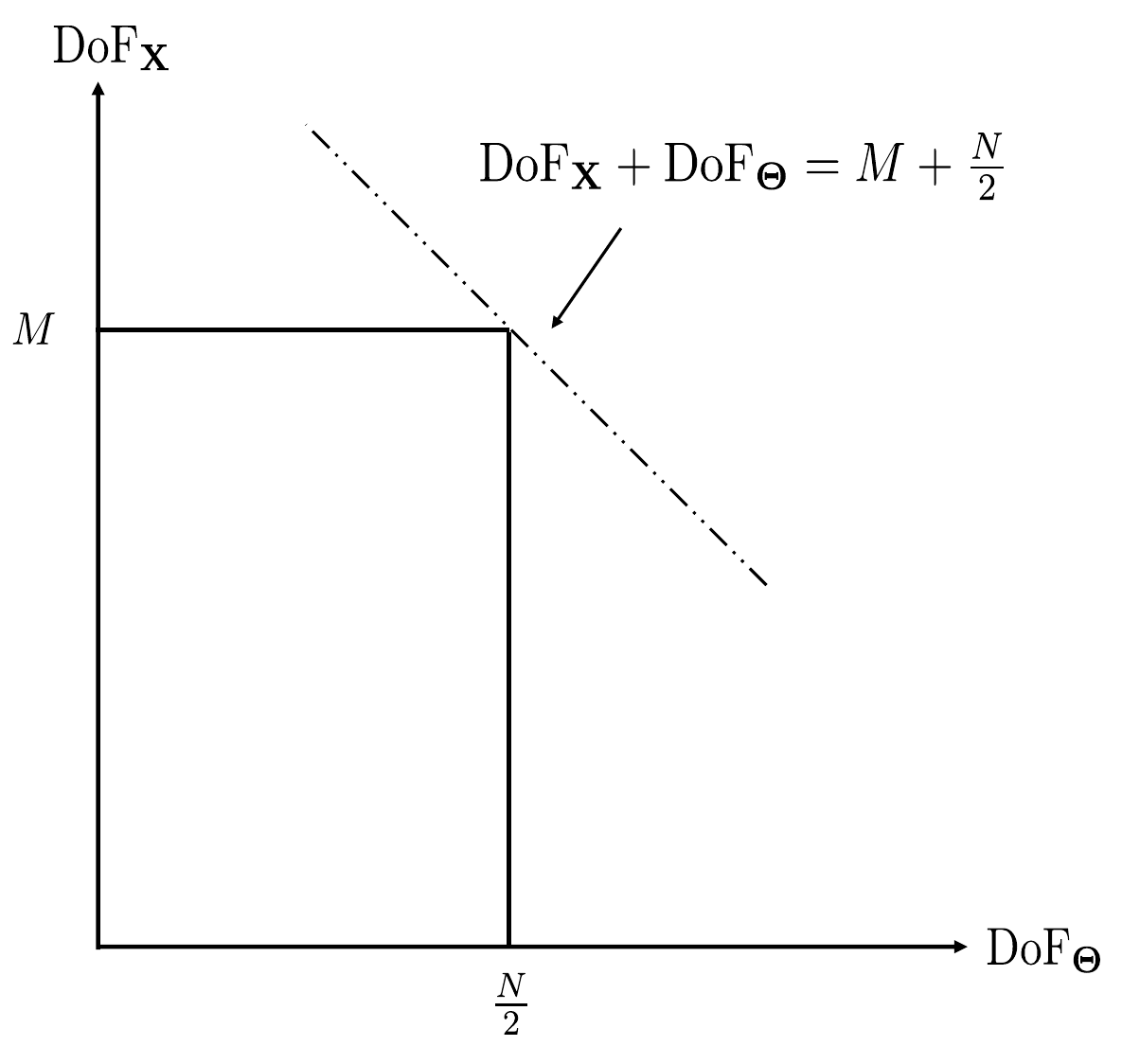}}
        \hspace{2.5mm}
	\subfigure[$M\leq K\leq \frac{N}{2}$ and $K\leq M+\frac{N}{2}$]{\includegraphics[height=0.2\textheight,keepaspectratio]{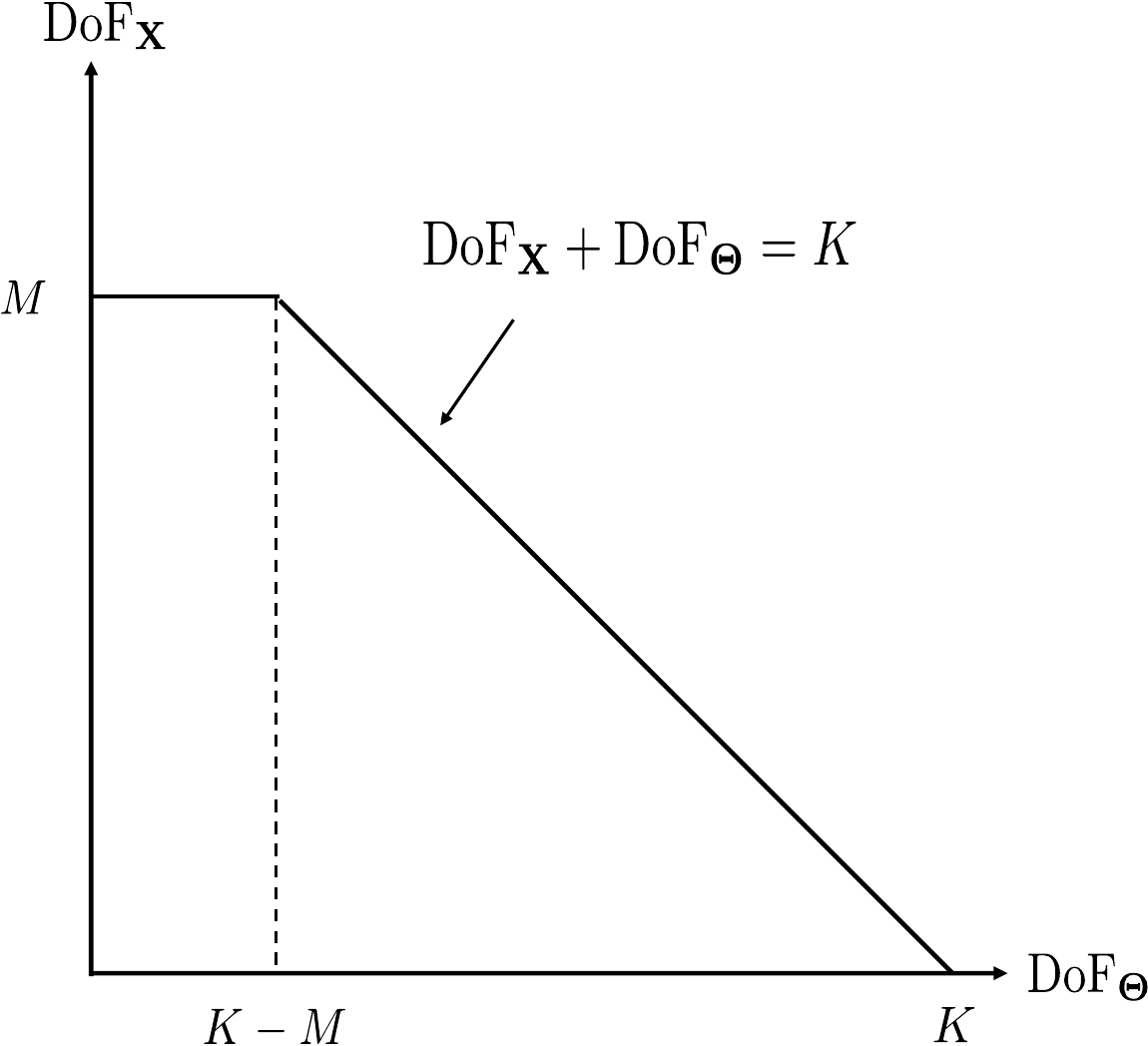}}
        \hspace{2.5mm}
	\subfigure[$\frac{N}{2} + r \le M \le N+r \le K $]{\includegraphics[height=0.2\textheight,keepaspectratio]{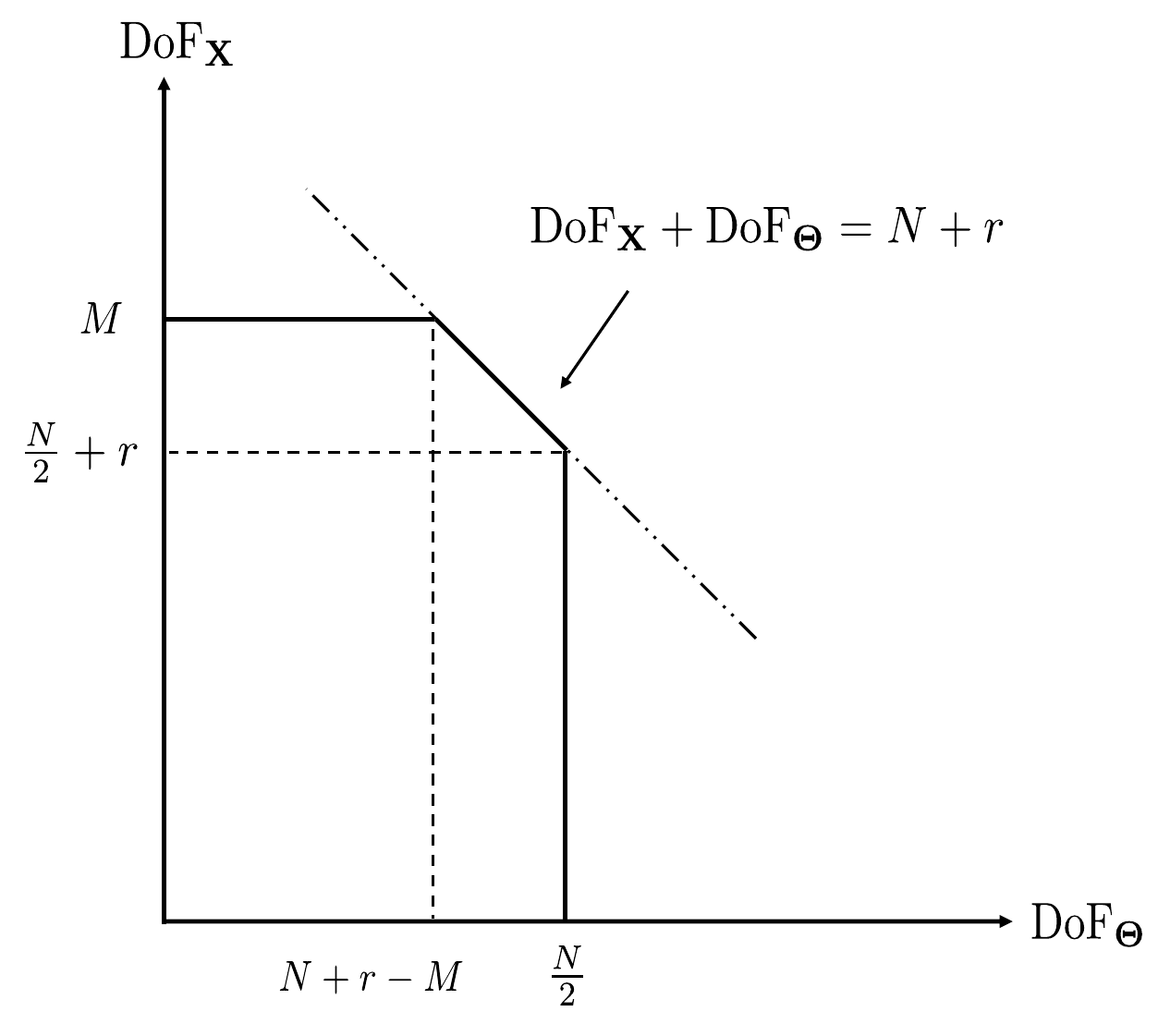}}
        \caption{Examples of DoF region for the multiple access channel with a direct path of rank $r$ under different values of $M$, $N$ and $K$.}
        \label{region3}
\end{figure*}

The DoF of the point-to-point channel for a channel model with a direct path
and with $(\mathbf{X},\mathbf{\Theta})$ as the input and $\mathbf{Y}$ as the
output is given by the following theorem.

\begin{theorem}\label{theorem_direct_path}
Consider the channel model 
\begin{equation}
\mathbf{Y} = \sqrt{P}\left(\mathbf{H} {\boldsymbol {\Theta}} \mathbf{G}+\mathbf{F}\right) \mathbf{X} + \mathbf{Z},
\end{equation}
with $(\mathbf{X}, \boldsymbol{\Theta})$ as the input where $\mathbf{X}\in \mathbb{C}^M$ has a power constraint $\mathbb{E}[\|\mathbf{X}\|_2^2]\leq 1$ and ${\boldsymbol{{\Theta}}}={\text{diag}}([e^{j\theta_1},e^{j\theta_2},\cdots,e^{j\theta_{{N}}}])$, $\mathbf{Y}\in\mathbb{C}^K$ as the output, and $\mathbf{Z} \sim\mathcal{CN}(0,\mathbf{I}_{K})$ as the noise. For almost all $\mathbf{G}$ and $\mathbf{H}$, and almost all direct path channels $\mathbf{F}$ with a fixed rank $r > 0$, the DoF from $(\mathbf{X}, \boldsymbol{\Theta})$ to $\mathbf{Y}$ is given by
\begin{equation}
\mathrm{DoF}_{(\mathbf{X}, \boldsymbol{\Theta})}=\min \left(M+\frac{N}{2}, N+r, K\right),
\end{equation}
which can be achieved with $\mathbf{X}$ and $\boldsymbol{\Theta}$ being independent.
\end{theorem}
\begin{IEEEproof}
Take the singular-value decomposition (SVD) of the rank-$r$ matrix $\mathbf{F}$
\begin{equation}
\mathbf{F}=\mathbf{U}\boldsymbol{\Sigma}\mathbf{V}, \label{eqn:svd}
\end{equation}
where $\mathbf{U}\in \mathcal{C}^{K\times r}$ and $\mathbf{V}\in \mathcal{C}^{r\times M}$ are unitary matrices, and $\boldsymbol{\Sigma}$ is a diagonal matrix of dimension $r$ with positive entries $\sigma_1, \ldots, \sigma_r$.
We rewrite the channel model as
\begin{align}
\mathbf{Y} &= \sqrt{P}\left(\mathbf{H} {\boldsymbol {\Theta}} \mathbf{G}+\mathbf{F}\right) \mathbf{X} + \mathbf{Z}, \\
&= \sqrt{P}\left(\mathbf{H} {\boldsymbol {\Theta}} \mathbf{G}+\mathbf{U}\boldsymbol{\Sigma}\mathbf{V}\right) \mathbf{X} + \mathbf{Z}, \\
&= \sqrt{P}[\mathbf{H}, \mathbf{U}\boldsymbol{\Sigma}] \left[
                                            \begin{array}{cc}
                                              {\boldsymbol {\Theta}} & \mathbf{0} \\
                                              \mathbf{0} & \mathbf{I}_r \\
                                            \end{array}
                                          \right]
\left[
  \begin{array}{c}
    \mathbf{G} \\
    \mathbf{V} \\
  \end{array}
\right] \mathbf{X} + \mathbf{Z}, \\
&= \sqrt{P}\mathbf{H'} \underbrace{{\boldsymbol {\Theta}'} \mathbf{G'} \mathbf{X}}_{\mathbf{W'}} + \mathbf{Z},
\label{systemmodel_3equ}
\end{align}
where $\mathbf{H'}=[\mathbf{H},\mathbf{U}\boldsymbol{\Sigma}]$, $\boldsymbol{\Theta'}=\left[
                                            \begin{array}{cc}
                                              {\boldsymbol {\Theta}} & \mathbf{0} \\
                                              \mathbf{0} & \mathbf{I}_r \\
                                            \end{array}
                                          \right] $ and $\mathbf{G'}=\left[
  \begin{array}{c}
    \mathbf{G} \\
    \mathbf{V} \\
  \end{array}
\right]$.
The channel model \eqref{systemmodel_3equ} is now a model without a direct path and with an RIS of effective size $N'=N+r$.
The intuition behind this reformulation is that the direct path channel can be regarded equivalently as a channel with an RIS of $r$ elements where the phase shifts are fixed at $0$ (or any arbitrary known phase).

Consider first the case with $r=1$. 
The overall system can now be viewed as a channel with an RIS of $N'=N+1$ reflective elements, but without a direct path. We choose the input distribution as $\mathbf{X}$ to be i.i.d.\ complex Gaussian for the $M$ elements, i.e., $\mathbf{X}\sim \mathcal{CN}(0,\mathbf{I}_{M})$,
and the first $N$ elements of $\boldsymbol{\Theta'}$ to follow uniform distribution with $\theta_i \sim \operatorname{Unif}[-\pi, \pi), ~i=1,\ldots, N$, , independent of $\mathbf X$. The last element of $\boldsymbol{\Theta'}$ is set to be fixed at $1$.
This choice of distributions is exactly the distributions in Case 2) of Theorem \ref{theoremNneqK}. From  Theorem \ref{theoremNneqK}, we see that the DoF achieved by this distribution for a system of size $(M, N',K)$ is given as
\begin{equation}
\min \left(M+\frac{N'}{2}-\frac{1}{2}, N',K \right)=\min\left(M+\frac{N}{2}, N+1, K\right).
\end{equation}
This proves the result for the case of $r=1$.

For the case of $r>1$, we need to make use of the techniques in the proof of Theorem \ref{theoremNneqK} involving characterizing the information dimension $D(\mathbf{W}')$ and its behavior under projection. The details are in Appendix \ref{app:proof_DoFdirect}.
\end{IEEEproof}

Theorem \ref{theorem_direct_path} shows that a direct path between the transmitter and the receiver helps resolve the global phase ambiguity and recover the $1/2$ DoF loss. Moreover, this can already be achieved with a direct path of channel rank of just $1$.

\subsection{Multiple Access Channel with $\mathbf{X}$ and $\mathbf{\Theta}$ as Inputs}
In this section, we consider the scenario in which the transmitter and the RIS
send independent messages to the receiver.
Similar to the case without the direct path, since the distribution that achieves the DoF in the joint transmission case as given in
Theorem \ref{theorem_direct_path} is already in the form of a product of two independent distributions, this gives the following
characterization of the DoF region of the multiple access channel with a direct path between the transmitter and the receiver.


\begin{theorem}\label{theorem_DoFdirect}
Consider the multiple access channel model
\begin{equation}
\mathbf{Y} = \sqrt{P}\left(\mathbf{H} {\boldsymbol {\Theta}} \mathbf{G}+\mathbf{F}\right) \mathbf{X} + \mathbf{Z},
\end{equation}
with $\mathbf{X}$ and $\boldsymbol{\Theta}$ as the two independent inputs, $\mathbf{Y}\in\mathbb{C}^K$ as the output, and $\mathbf{Z} \sim\mathcal{CN}(0,\mathbf{I}_{K})$ as the noise. Here $\mathbf{X}\in \mathbb{C}^M$ has a power constraint $\mathbb{E}[\|\mathbf{X}\|_2^2]\leq 1$ and ${\boldsymbol{{\Theta}}}={\text{diag}}([e^{j\theta_1},e^{j\theta_2},\cdots,e^{j\theta_{{N}}}])$.
For almost all $\mathbf{G}$ and $\mathbf{H}$, and almost all direct path channel $\mathbf{F}$ with a fixed rank $r>0$, the DoF region of the multiple access channel is given by the set of DoF pairs ($\mathrm{DoF}_{\mathbf{X}}$,$\mathrm{DoF}_{\boldsymbol{\Theta}}$) that satisfy
\begin{eqnarray}
\mathrm{DoF}_{\mathbf{X}} & \leq & \min(M, N+r, K), \label{eq:mac_direct_1} \\
\mathrm{DoF}_{\boldsymbol{\Theta}} & \leq & \min\left(\frac{N}{2},K\right), \label{eq:mac_direct_2} \\
\mathrm{DoF}_{\mathbf{X}}+\mathrm{DoF}_{\boldsymbol{\Theta}}  & \leq & \min \left(M+ \frac{N}{2}, N+r, K \right). \label{eq:mac_direct_3}
\end{eqnarray}
\end{theorem}
\begin{IEEEproof}
First, we show the converse. We rewrite the model as $\mathbf{Y}=\sqrt{P}\mathbf{H'} \boldsymbol {\Theta}' \mathbf{G'} \mathbf{X} + \mathbf{Z}$ as in the proof of Theorem \ref{theorem_direct_path}. 
To bound $I(\mathbf{X};\mathbf{Y}|\mathbf{\Theta})$, observe that conditioned on $\mathbf{\Theta}$, the channel from $\mathbf{X}$ to $\mathbf{Y}$ is equivalent to a MIMO channel with a channel matrix $\mathbf{H}_{\mathrm{eq}}=\mathbf{H}\boldsymbol{\Theta}'\mathbf{G}'$.
Thus, $\mathrm{DoF}_{\mathbf{X}}$ is bounded by the rank of the equivalent channel as follows: 
\begin{equation}
\mathrm{DoF}_{\mathbf{X}} \leq \mathrm{rank} (\mathbf{H}_{\mathrm{eq}}) =\min (M,N+r,K).
\end{equation}
Thus, an upper bound of $\mathrm{DoF}_{\mathbf{X}}$ is given by (\ref{eq:mac_direct_1}).
Similarly, $\mathrm{DoF}_{\boldsymbol{\Theta}}$ is upper bounded by the input dimension $\frac{N}{2}$ and the output dimension $K$, which give (\ref{eq:mac_direct_2}).
Finally, the sum DoF is upper bounded by allowing cooperation between $\mathbf{X}$ and $\boldsymbol{\Theta}$. By Theorem \ref{theorem_direct_path}, we have
(\ref{eq:mac_direct_3}).

Next, we show achievability. Recognizing that the sum DoF as characterized in Theorem \ref{theorem_direct_path} is achieved with an independent distribution $p(\mathbf{X})p(\mathbf{\Theta})$, it must also be an achievable sum DoF in the multiple access setting.
By choosing to decode $\mathbf{X}$ first, then $\mathbf{\Theta}$,
we get the following achievable DoF pair corresponding to one corner point
of the DoF region outer bound:
\begin{align}
& \mathrm{DoF}_{\boldsymbol{\Theta}} = \min\left(\frac{N}{2},K\right),\\
& \mathrm{DoF}_{\mathbf{X}} =
 \min \left(M+ \frac{N}{2}, N+r, K \right) - \min\left(\frac{N}{2},K\right).
\end{align}
The opposite decoding order achieves the other corner point.
This shows that the entire region (\ref{eq:mac_direct_1})-(\ref{eq:mac_direct_3}) is achievable.
\end{IEEEproof}

As the case without the direct path, depending on the relative values of $M$,
$N$ and $K$, the DoF region can take different shapes as shown by the three
examples in Fig.~\ref{region3}. In all three examples, modulating information
through RIS allows additional data streams to be transmitted to the receiver.
The effect of the direct path is evident in Case (a), where the direct path
helps resolve the phase ambiguity hence removing the $1/2$ dimension overlap,
and in Case (c), where the sum DoF is improved by the rank of the direct path.

As an additional remark, similar to the case without the direct path, 
this result can be generalized to the scenarios in which $\mathbf{X}$ or
$\boldsymbol{\Theta}$ can consist of multiple users, i.e., different
components of $\mathbf{X}$ and $\mathbf{\Theta}$ can represent different users or different RISs, respectively.


\section{Operating RIS at Different Rate as Transmitter}\label{sec:different_rate}

The discrete-time channel model considered so far in the paper implicitly
assumes that the symbol rate of $\mathbf{X}$ is the same as the rate of phase shifts
$\boldsymbol{\Theta}$ at the RIS. In this section, we look at a scenario in which the
phase shifts of the RIS operate at a slower rate---specifically, it must remain constant
for $n$ symbols of $\mathbf{X}$. In this case, the channel model without a direct path is as follows:
\begin{equation}
\mathbf{Y}(t) = \sqrt{P}\mathbf{H} {\boldsymbol {\Theta}} \mathbf{G} \mathbf{X}(t) + \mathbf{Z}(t), \qquad t=1,\ldots,n, \label{systemmodel_2}
\end{equation}
where over $n$ time slots, $\mathbf{X}(t)$ can take $n$ different symbols, but
$\boldsymbol {\Theta}$ must remain constant. For simplicity, we assume that the channels
$\mathbf{H}$ and $\mathbf{G}$ are fixed over the $n$ time slots.
Gathering the received signals from the $n$ time slots and writing it in a vector form, we have
\begin{equation}
\mathbf{\check{Y}}= \sqrt{P}\left(\mathbf{I}_n \otimes \mathbf{H} {\boldsymbol {\Theta}}\mathbf{G}\right) \mathbf{\check{X}}  + \mathbf{\check{Z}}, \label{systemmodel_matrix}
\end{equation}
where $\mathbf{\check{Y}}=[\mathbf{Y}(1)^\mathsf{T},\ldots,\mathbf{Y}(n)^\mathsf{T}]^\mathsf{T}$,
$\mathbf{\check{X}}=[\mathbf{X}(1)^\mathsf{T},\ldots,\mathbf{X}(n)^\mathsf{T}]^\mathsf{T}$,
and $\mathbf{\check{Z}}=[\mathbf{Z}(1)^\mathsf{T},\ldots,\mathbf{Z}(n)^\mathsf{T}]^\mathsf{T}$,
are the received signals, transmit signals and noises, respectively, across the $n$ time slots.
Here, $(\cdot)^\mathsf{T}$ denotes transpose.

Using the same proof technique as in Theorem \ref{theoremNneqK}, we have the following DoF result:
\begin{theorem}
For the channel model \eqref{systemmodel_2} with $\left(\mathbf{{X}}(1),\cdots,\mathbf{X}(n), \mathbf{\Theta}\right)$ as the input and $(\mathbf{Y}(1),\cdots,\mathbf{Y}(n))$ as output,
under power constraint $\mathbb{E}[\| \mathbf{X}(t) \|_2^2] \le 1, \forall t$,
for almost all matrices $\mathbf{H}$ and $\mathbf{G}$,
the DoF of the overall channel over $n$ samples is given by
\begin{equation}
\mathrm{DoF}_n=\min\left(nM+\frac{N-1}{2}, nN, nK\right).
\end{equation}
\end{theorem}
%
This result can be extended to the multiple-access case and to the case with direct path. 

\section{Modulating Information Through Phases}
\label{sec:slp}

The DoF analysis of the previous sections gives theoretic results on how many
independent data streams can be transmitted through the phases of an RIS system.
In this section, we suggest two different ways of modulating information
through the phases of the RIS.

First, we describe
a maximum likelihood (ML) decoding strategy to jointly decode the modulated information in $\mathbf{X}$ and $\boldsymbol{\Theta}$. Second, we propose a symbol-level precoding scheme to jointly encode information in $\mathbf{X}$ and $\boldsymbol{\Theta}$.

\subsection{Maximum Likelihood Decoding}

A straightforward way to modulate information jointly through the transmitter
and the phases of the RIS is to use fixed constellations to encode the
information at $\mathbf{X}$ and $\boldsymbol \Theta$, then use maximum
likelihood (ML) decoding to jointly decode the transmitted symbols.
Due to the phase-only nature of RIS, PSK constellation can be used at
$\boldsymbol{\Theta}$. The encoding at $\mathbf X$ can be in the form of
symbols from constellation points multiplied by a fixed precoding vector
(or the precoding matrix in case of multiple data streams).
Assuming the general case of an RIS system with a direct path,
the ML estimates of the transmitted symbols at the decoder are given by:
\begin{equation}\label{ML}
({\mathbf{\hat X}}, {\boldsymbol {\hat\Theta}}) =
\mathrm{argmin}_{(\mathbf{X},\boldsymbol {\Theta}) \in {\mathcal C}} {\|\mathbf{Y}-\sqrt{P}(\mathbf{H}\boldsymbol {\Theta}\mathbf{G}+\mathbf{F})\mathbf{X}\|_2^2}.
\end{equation}
where the minimization is over the constellations of $\mathbf X$ and $\boldsymbol \Theta$, denoted as $\mathcal{C}$ in the above. This is a discrete
optimization problem for which finding a global optimal solution would have
a high computational complexity.
This approach of modulating information in the phase shifts at RIS has been explored in \cite{yan2020} and \cite{reflecting_modulation} by solving problem \eqref{ML} using approximation algorithms.


\subsection{Symbol-Level Precoding}

As an alternative to encoding information using fixed constellations at
the transmitter $\mathbf X$ and the RIS $\boldsymbol \Theta$, in this section
we propose a symbol-level precoding strategy that sets the received signal
$\mathbf Y$ to be as close to the desired points in a fixed constellation
as possible.  The idea is to try to directly make $\mathbf Y$ to be the
symbol corresponding to the information to be transmitted, and to design
$\mathbf X$ and $\boldsymbol \Theta$ so that the desired symbols are
synthesized at the receiver. Because $\mathbf X$ and $\boldsymbol \Theta$ are
jointly designed, this strategy is applicable to the joint transmission case,
but not to the multiple access setting.

In the below, we discuss the symbol-level precoding problem formulation
and the numerical algorithms for solving the precoding problem.


\subsubsection{Problem Formulation}

Given a vector ${\mathbf{\hat Y}}$ corresponding to the desired
constellation points at the receiver and assuming perfect CSI, symbol-level
precoding proceeds by solving the following optimization problem:
%
\begin{equation}\label{slp2}
	({\mathbf{X^*}}, {\boldsymbol {\Theta^*}})=\mathrm{argmin}_{\mathbf{X}, \boldsymbol{\Theta}} {\|\hat{\mathbf{Y}}-\sqrt{P}\left(\mathbf{H}\boldsymbol {\Theta}\mathbf{G}+\mathbf{F}\right)\mathbf{X}\|_2^2}.
\end{equation}
subject to that $\|\mathbf{X}\|^2_2 \le 1$ and
$\boldsymbol{\Theta} = {\rm diag}([e^{j\theta_1}, \cdots, e^{j\theta_N}])$.
%
A special case of this is when the transmit signal $\mathbf{X}$ is deterministic as studied in Section \ref{sec:knownX}, for which the optimization problem can be formulated as:
\begin{equation}\label{slp3}
	{\boldsymbol{\Phi^*}}=\mathrm{argmin}_{\boldsymbol {\Phi}} \|\bar{\mathbf{Y}}- \sqrt{P} \bar{\mathbf{H}}\mathbf{\Phi}\|_2^2,
\end{equation}
where $\boldsymbol{\Phi}=[e^{j\theta_1},\ldots,e^{j\theta_{N}}]^{T}$ is the vector form of the diagonal matrix $\boldsymbol{\Theta}$,
$\bar{\mathbf{H}}=\mathbf{H}\mathrm{diag}(\mathbf{G}\mathbf{X})$ is the equivalent channel, 
and $\bar{\mathbf{Y}}=\mathbf{\hat Y}-\mathbf{FX}$ is the effective received signal. This special case has the same formulation as the problem of MIMO precoding with per-antenna constant envelope constraints, as studied in \cite{constant_envelope2013,maConstantEnvelope2014}.

If the above optimization problem can be solved to achieve a (near)-zero minimum, the resulting (${\mathbf{X^*}}, {\boldsymbol {\Theta^*}}$) can be directly used as the transmit signal and the reflective coefficients, in which case the desired constellation points plus noise would be received at $\mathbf{Y}$. 

This symbol-level precoding strategy has been proposed earlier for MIMO channels
in various formulations \cite{slp_ottersten2015,slp_masouros2015}.
In this paper, we apply the concept to the RIS system. Moreover, we point out
that it is possible to numerically evaluate the relationships between the
dimensions of $\mathbf X$, $\mathbf \Theta$ and $\mathbf Y$ that allow this
optimization problem to be solved to a near-zero minimum. This gives a method
to numerically characterize the achievable DoF of the RIS system.

\subsubsection{Numerical Algorithm}


A main advantage of the symbol-level precoding formulation is that instead of
solving a discrete optimization problem (\ref{ML}), we now solve a continuous
optimization problem (\ref{slp2}). Although (\ref{slp2}) is not convex,
as it involves a product of optimization variables in a bilinear form as the
objective and unit modulus constraints, there are optimization techniques
available that allow efficient gradient search to find local optimal solutions.



Specifically, this paper proposes an approach based on the augmented Lagrangian approach \cite{alm_boumal2020} to take care of the power constraint on $\mathbf{X}$ and a Riemannian conjugate gradient method \cite{manifold_IRS2019} to account for the unit modulus constraint on $\boldsymbol{\Theta}$ to solve (\ref{slp2}).
The Riemannian conjugate gradient method recognizes
the fact that the unit modulus constraints force
$\mathbf{\Theta}$ to be on a \textit{complex circle manifold}, and modifies the
traditional gradient descent algorithm (which works on the Euclidean space) to work on the
manifold to enhance the convergence speed. 
A description of the numerical method
for solving the optimization problems \eqref{slp2} is given in
Appendix \ref{app:riemannian}.  This algorithm is used in the next
section to numerically verify the DoF analysis for modulating information
through the phases of the RIS.

%

\section{Numerical Results}
\label{sec:sim}

In this section, we present numerical results based on the symbol-level
precoding approach to validate the DoF analysis of jointly modulating
information by the transmitter and the RIS for the receiver.

\subsection{Channel Model and Simulation Setting}

We consider a MIMO communication system with an RIS, consisting of a transmitter with $2$ antennas and a receiver with 4 antennas. The number of elements of the RIS varies in different simulation settings. All the channel entries follow i.i.d.\ Rayleigh fading with unit variance, i.e., $\mathcal{CN}(0,1)$.

The idea is to perform symbol-level precoding to synthesize the desired
constellation point $\mathbf{Y}$ by solving for $\mathbf{X}$ and
$\boldsymbol{\Theta}$ in (\ref{slp2}) or (\ref{slp3}). We use the Riemannian
conjugate gradient approach to find a local optimal solution,
and consider a point to be feasible if the synthesized point is within a
distance threshold of $\delta=10^{-3}$ from the desired point.


\begin{figure*}
    \centering
    \subfigure[$N=4$]{\includegraphics[width=0.32\textwidth]{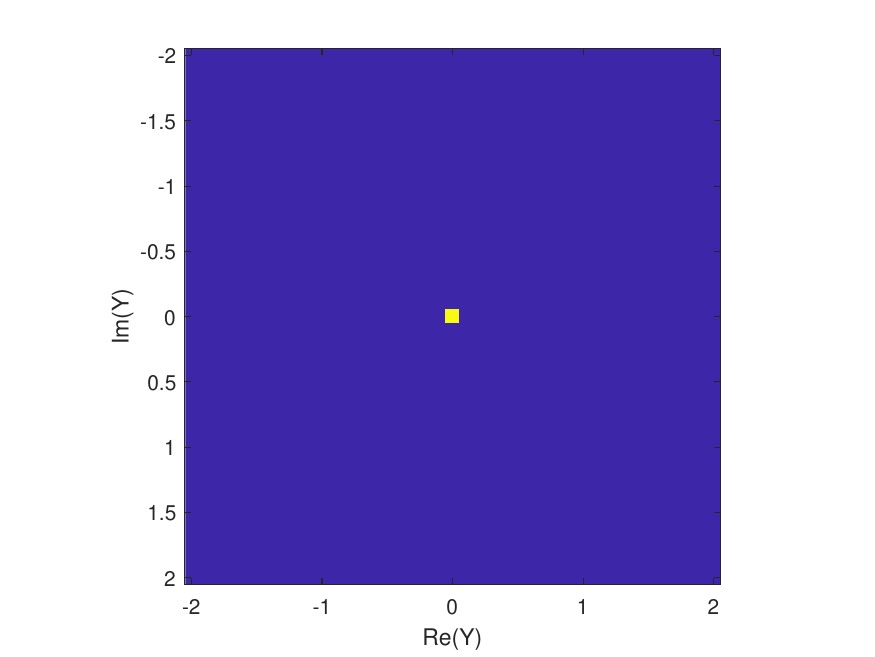}}
    \subfigure[$N=5$]{\includegraphics[width=0.32\textwidth]{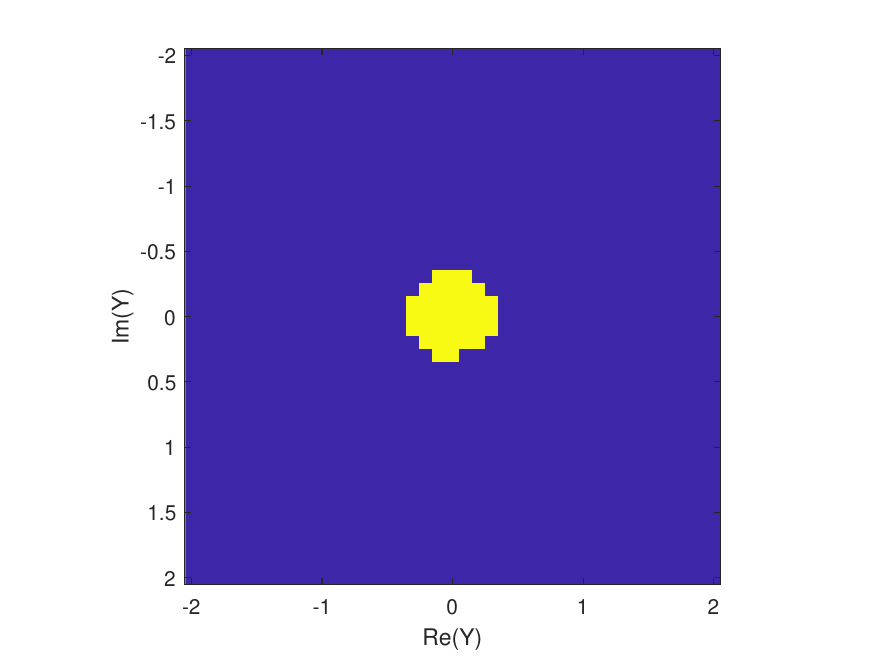}}
    \subfigure[$N=6$]{\includegraphics[width=0.32\textwidth]{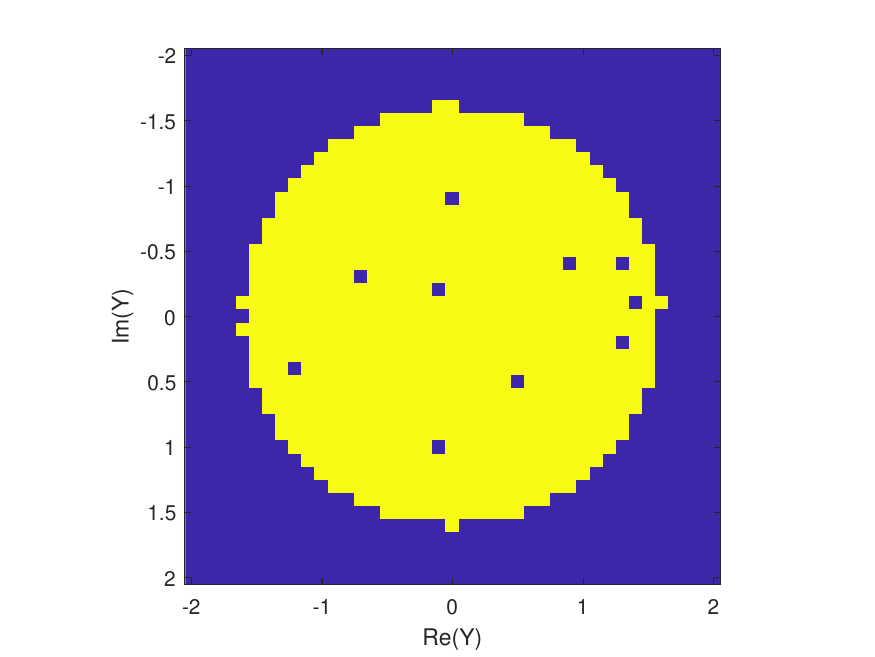}}
	\caption{The feasible region of $y$ for synthesizing $\mathbf{Y}=[y,y,y,y]$ at $M=2$, $K=4$, $P=1$, and $N$ ranging from 4 to 6 for channels without the direct path. Color yellow means that a point is feasible while color blue means that a point is infeasible.} 
    \label{fig:feasible_region}
\end{figure*}

\begin{figure}
        \centering
        \includegraphics[width=0.5\textwidth]{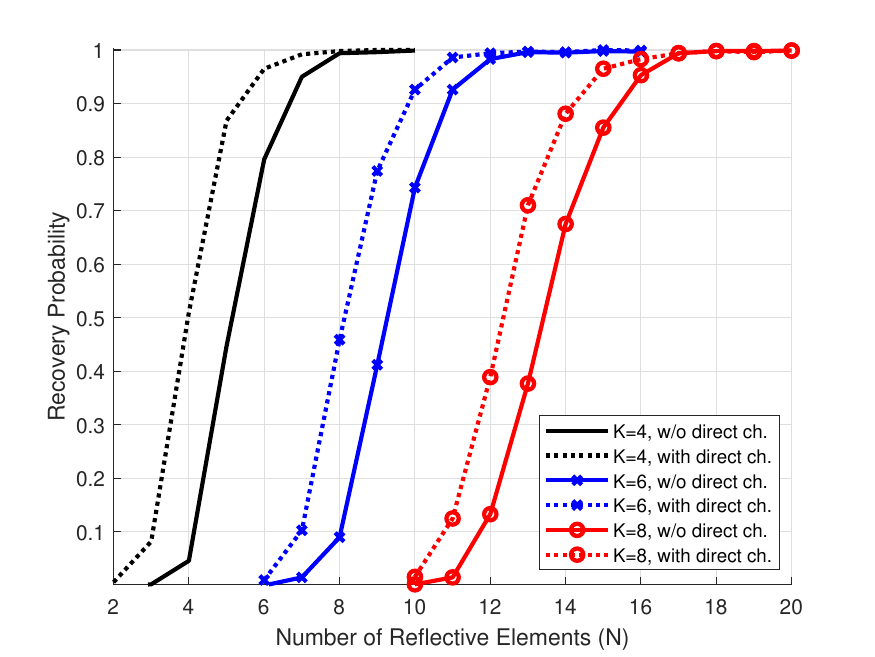}
        \caption{Probability of successfully synthesizing $\mathbf{Y}$ at the receiver with each component of $\mathbf{Y}$ a random point on the unit circle 
	for different values of $N$, with $M=2$ and $P=10$.}

        \label{recovering_1}
\end{figure}

\begin{figure}
        \centering
        \includegraphics[width=0.5\textwidth]{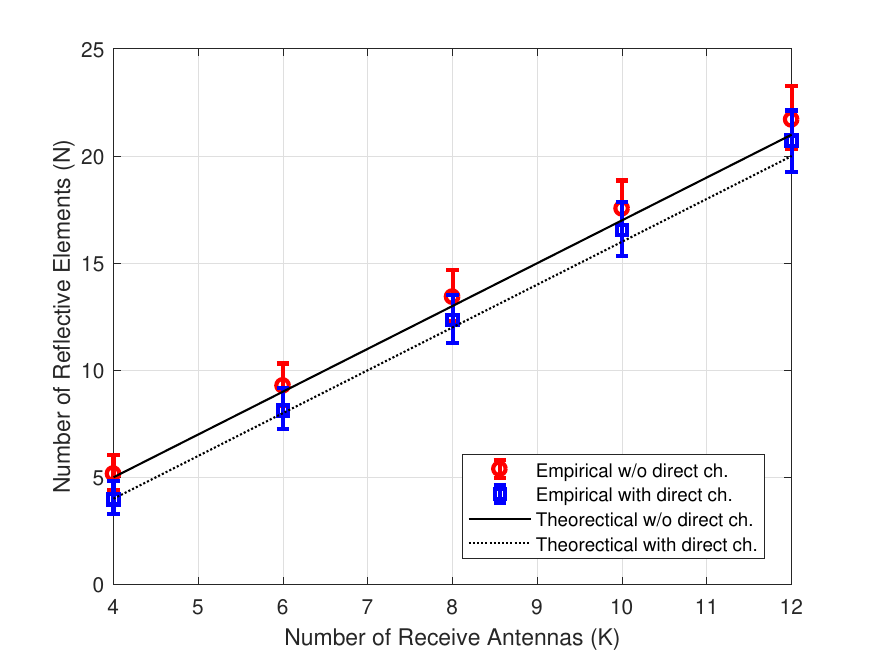}
	\caption{The number of RIS elements $N$ needed to successfully synthesize a $K$-dimensional $\mathbf{Y}$ with each element of $\mathbf{Y}$ being a random point on the complex unit circle. The theoretical lines correspond to the DoF results in Theorems \ref{theoremNneqK} and \ref{theorem_direct_path}, i.e., $N=2K-2M$ and $N=2K-2M+1$ for the case with and without the direct path, respectively.
The error bars of the empirical points correspond to 80, 50, and 20-percentile probabilities of success for synthesizing $\mathbf{Y}$.}
        \label{errorbar}
\end{figure}

\subsection{Dimension of the Symbol-Level Precoded Signal}

The DoF of joint transmission through the transmitter and the RIS can be
numerically verified by visualizing the dimension of the set of feasible
receive points $\mathbf Y$. For example, in the case with no direct path,
assuming that $M+\frac{N}{2}-\frac{1}{2}\geq K$,
the set of points that $\mathbf{H\Theta G}\mathbf{X}$ can reach should form
a $K$-dimensional set in the complex space. In other words, for each element of
$\mathbf{Y}$, the set of reachable points by designing $\mathbf X$ and
$\boldsymbol \Theta$ should be a two-dimensional set in the real and imaginary
plane.

Fig.~\ref{fig:feasible_region} shows an example of this phenomenon with $M=2$,
$K=4$ and $N$ varying from $4$ to $6$ for the case of without the direct path.
The transmit power $P$ is set to be $1$. The feasible regions for $y$ with
$\mathbf{Y}=[y, y, y, y]^T$ are shown with color marked to represent the
feasibility of the points. Yellow means that a point is feasible;
blue means that a point is infeasible. A first observation is that
increasing the effective transmit dimension $M+\frac{N}{2}-\frac{1}{2}$
enlarges the feasible region. At $N=4$, no points are feasible except the origin. When
$N=5$, the feasible region becomes a two-dimensional region in each component
of $\mathbf{Y}$. At $N=6$, the feasibility region is further enlarged.
This agrees with the theoretical analysis that $N=5$ is needed to achieve a DoF of $4$.

\subsection{Phase Transition for Reaching the Desired Points}

To further illustrate the phase transition from infeasibility to
feasibility, we plot the probability of reaching a desired point in
Fig.~\ref{recovering_1}.
We choose $y_1,\ldots, y_K$ to be random points on the complex unit circle and aim to synthesize $\mathbf{Y}=[y_1,\ldots,y_K]^T$ as the desired point. We increase the transmit power to $P=10$ in order to make the desired points to be in the feasible region (should the region exist). This is to avoid the situation in which the infeasibility is due to the transmit power being too low. 

Fig.~\ref{recovering_1} shows the probability of successful synthesis against
the number of elements at the RIS. It illustrates the phase transition with
respect to
$N$. The number of transmit antennas is $M=2$.  The cases for $K=4,~6,~8$ are
plotted on the same scale.  Each plot is generated by solving \eqref{slp2} for the scenarios with or without the direct path, respectively, over
1000 channel realizations.
From Fig.~\ref{recovering_1}, we observe that the phase
transition has a median at around $N=2K-2M+1$ for the case of without the direct path,
and $N=2K-2M$ with the direct path, which verifies the DoF results. Further, there is a gap of one RIS element
between the two cases, which corresponds to the
$1/2$ dimension difference in the DoF result. This matches
Theorem \ref{theoremNneqK} and Theorem \ref{theorem_DoFdirect}.








To further illustrate the phase transition,
Fig.~\ref{errorbar} shows the number of RIS elements corresponding to the $20$, $50$ and $80$ percentile probabilities of successfully synthesizing a received vector $\mathbf{Y}$ with $K$ antennas. The theoretical DoF curves of
$N=2K-2M$ and $N=2K-2M+1$ are also plotted, for the scenarios with and without the direct path, respectively.
From Fig.~\ref{errorbar}, we observe that the theoretical DoF curves are slightly below the $50$-percentile points.
The gap between the $80$-percentile points and the theoretical
DoF curves is about $1$ to $2$ RIS elements, and the gap between the
$20$-percentile points and the theoretical DoF curves is about $0$ to $1$ RIS
element. These results verify the theoretical analysis. It also shows that to
ensure consistent feasibility of synthesizing a received vector, it is important to have
slightly more than the minimum number of RIS elements.


\section{Conclusions}
\label{sec:conclusion}

This paper studies the information theoretic limits of joint information
transmission using both signaling at the input and phase modulation at the RIS reflector.
In particular, the DoFs (or the multiplexing gains) of the communication schemes where
information is conveyed through the transmitted symbols and the reflective
coefficients of the RIS are characterized for both the point-to-point and the
multiple access settings. This proposed use of the RIS significantly improves
the DoF of the overall system as compared to the
conventional paradigm of simply using the RIS as a passive beamformer for
enhancing the channel strength.

The theoretical DoF analysis in this paper is obtained
by recognizing a connection between the RIS channel and the MIMO channel with phase noise. Tools
for studying the information dimension and the point-wise dimension under
projection are used to obtain the DoF result. For example, as compared to
a MIMO channel with $M$ transmit antennas and $K$ receive antennas, which
has a DoF of $\min(M,K)$, an RIS system with the same number of transmit and
receive antennas and further equipped with a reconfigurable surface of
$N$ reflective elements can achieve an overall DoF of
$\min(M+\frac{N}{2}-\frac{1}{2},N,K)$ by allowing joint encoding by the
transmitter and the RIS, (assuming no direct path between the transmitter and
the receiver).  This shows the significant potential of using the RIS to
modulate information and to improve the overall transmission rate in a system
limited by the number of transmit antennas.

Finally, this paper proposes a symbol-level precoding strategy to enable
the practical realization of modulating information through the phases of the
RIS. We show numerically that the phase transition for successful symbol-level
precoding agrees with the theoretical DoF analysis.



\appendices

\section{Information Dimension Under Projection and \\ Proof of Theorem \ref{theoremNneqK}}
\label{app:proof_DoFp2p}

We present the proof of Theorem \ref{theoremNneqK} by first building a connection between the model \eqref{systemmodel} and the MIMO channel model with phase noise. Then, we make use of the concepts of information dimension and point-wise dimension (see \cite{Wu2015,Hunt1997}), and study their behavior under projections, to compute the DoF of \eqref{systemmodel}. 

\subsection{Information Dimension and DoF}


\begin{definition}\label{alternative}
For a random vector $\mathbf{X} \in \mathbb{R}^N$ with distribution $\mu$, we define the lower and upper information dimension of $\mathbf{X}$ as
\begin{align}	
&\underline{D}(\mathbf{X})=\liminf_{\varepsilon\to 0}\frac{\mathbb {E} \left[ \log \mu \left(B(\mathbf{X};\varepsilon )\right )\right ]}{\log \varepsilon}	
\end{align}
\text{and}
\begin{align}
&\overline{D}(\mathbf{X})= \limsup_{\varepsilon\to 0}\frac{\mathbb {E} \left[ \log \mu \left(B(\mathbf{X};\varepsilon )\right )\right ]}{\log \varepsilon},
\end{align}
respectively, where $B(\mathbf{X};\varepsilon)\subseteq \mathbb R^N$ denotes the ball with center $\mathbf{X}$ and radius $\varepsilon$ with respect to an arbitrary  norm on $\mathbb R^N$. When $\underline{D}(\mathbf{X})=\overline{D}(\mathbf{X}) = D(\mathbf{X})$, information dimension of $\mathbf{X}$ exists and is defined as $D(\mathbf{X})$.
\end{definition}

The following lemma connects the concept of information dimension to the multiplexing gain of a general vector additive noise channel.

\begin{lemma}[\cite{Wu2015},\cite{Stotz2016}]\label{connection}
Let $\mathbf{X}$ and $\mathbf{Z}$ be  independent random vectors in $\mathbb R^n$ such that $\mathbf{Z}$ has an absolutely continuous distribution with $h(\mathbf{Z})>-\infty$ and $H(\lfloor \mathbf{Z}\rfloor )<\infty$. Then
\begin{align}	
\limsup_{P\to\infty}\frac{I(\mathbf{X};\sqrt{P} \mathbf{X}+\mathbf{Z})}{\frac{1}{2}\log P}=D(\mathbf{X}).	
\end{align}
\end{lemma}

By Lemma \ref{connection}, we see that computing the multiplexing gain of the RIS channel \eqref{systemmodel} can be equivalently cast as computing the information dimension $D(\mathbf{H}\boldsymbol{\Theta}\mathbf{GX})$.
However, Lemma \ref{connection} deals with real channels. To work with a complex channel model, we can stack the real and imaginary parts of the complex vectors, and stack the real and imaginary parts of complex matrices, such as $\mathbf{H}$, as
\begin{align}
\left[
        \begin{array}{cc}
		\Re\{\mathbf{H}\} & -\Im\{\mathbf{H}\} \\
		\Im\{\mathbf{H}\} & \Re\{\mathbf{H}\} \\
        \end{array}
\right],
\end{align}
then apply Lemma \ref{connection} to the resulting equivalent model in $\mathbb{R}^{2K}$.
Note that two real dimensions in $\mathbb{R}^{2K}$ is equivalent to one complex dimension in  $\mathbb{C}^{K}$.

\subsection{Information Dimension Under Projection}

Our goal is to characterize the DoF of the channel model (\ref{systemmodel}). We do this by investigating the information dimension of $\mathbf{HW}$.
Observe that $D(\mathbf{HW})$ is the information dimension of $\mathbf{W}$ under a linear projection from $\mathbb C^N \rightarrow \mathbb C^K$, so one would expect that
\begin{equation}
D(\mathbf{HW}) = \min( D(\mathbf{W}), K).
\label{eq:min_project}
\end{equation}
The above relationship is, however, not true for any arbitrary distribution; see \cite{Wu2015,Hunt1997} for a counterexample. The issue is that the projection may have different effects on different points if the distribution is not absolutely continuous.
For the particular $\mathbf W$ studied in the context of this paper, the
relationship (\ref{eq:min_project}) turns out to be true. To establish this
rigorously, we formally state several results on information dimension under
projection in this section. First, we need to define the concept of point-wise
dimension.


\begin{definition}\label{pointwise}
For every point $\mathbf{x}$ and a probability distribution $\mu$, we define the lower and upper point-wise dimension at $\mathbf{x}$ on the support of $\mu$ as
\begin{align}	
&\underline{d}(\mathbf{x})=\liminf_{\varepsilon\to 0}\frac{ \log \mu \left(B(\mathbf{x};\varepsilon )\right )}{\log \varepsilon}	 
\end{align}
\text{and}
\begin{align}
&\overline{d}(\mathbf{x})= \limsup_{\varepsilon\to 0}\frac{ \log \mu \left(B(\mathbf{x};\varepsilon )\right )}{\log \varepsilon},
\end{align}
respectively.
When $\underline{d}(\mathbf{x})=\overline{d}(\mathbf{x}) = d(\mathbf{x})$, point-wise dimension exists at $\mathbf{x}$ and is defined as $d(\mathbf{x})$.
\end{definition}

Next we present a lemma that provides an upper bound for the information dimension and point-wise dimension of a random vector $\mathbf{X}$ under a Lipschitz continuous mapping, which is later used to establish the converse of the DoF results. This lemma is a modification of the first part of Theorem 4.1 and the first part of Corollary 4.2 in \cite{Hunt1997}, but with the assumption for compact support removed and the condition of Lipschitz continuity added.

\begin{lemma}\label{projection_ub}
For any Lipschitz continuous mapping $f: \mathbb{R}^S \rightarrow \mathbb R^T$ with Lipschitz constant $\lambda$ and $\mathbf{X} \in \mathbb R^S$ with probability measure $\mu$, the following inequalities for information dimension always hold:
\begin{align}
&\overline{D}(f(\mathbf{X})) \leq \min(\overline{D}(\mathbf{X}),T), \\
&\underline{D}(f(\mathbf{X})) \leq \min(\underline{D}(\mathbf{X}),T).
\end{align}
Similarly, the following inequalities for point-wise dimension hold for all $x$:
\begin{align}
&\overline{d}(f(\mathbf{x})) \leq \min(\overline{d}(\mathbf{x}),T), \\
&\underline{d}(f(\mathbf{x})) \leq \min(\underline{d}(\mathbf{x}),T).
\end{align}
\end{lemma}

\begin{IEEEproof}
The proof is based on the proof of Theorem 4.1 in \cite{Hunt1997}, but instead of requiring the mapping to have compact support, we require the mapping to be Lipschitz continuous. Given that $f$ is Lipschitz continuous, it is continuous and thus measurable. From the Lipschitz continuous condition, we have
\begin{equation}
B\left(\mathbf{x};\frac{\varepsilon}{\lambda}\right)\subseteq f^{-1}(B(f(\mathbf{x}),\varepsilon)),
\end{equation}
where $f^{-1}(A)$ of a set $A$ is defined to be the set $\{a\in \mathbb{R}^S|f(a)\in A\}$.
Define the push forward measure to be $f(\mu)(A)=\mu(f^{-1}(A))$ for a set A.
This implies
\begin{equation}\label{measure}
f(\mu)(B(f(\mathbf{x}),\varepsilon))=\mu(f^{-1}(B(f(\mathbf{x}),\varepsilon))\geq \mu\left(B\left(\mathbf{x};\frac{\varepsilon}{\lambda}\right)\right).
\end{equation}
Dividing both sides by $\log \varepsilon$ and take $\limsup$ yields
\begin{equation}
\overline{d}(f(\mathbf{x})) \leq \min(\overline{d}(\mathbf{d}),T)
\end{equation}
for all $\mathbf{x}$.

Taking expectation on both sides of \eqref{measure} with respect to $\mu$, dividing both size by $\log \varepsilon$ and take $\limsup$ yields
\begin{equation}
\overline{D}(f(\mathbf{X})) \leq \min(\overline{D}(\mathbf{X}),T).
\end{equation}
The same line of proof works also for $\lim \inf$. Thus,
\begin{equation}
\underline{d}(f(\mathbf{x})) \leq \min(\underline{d}(\mathbf{x}),T)
\end{equation}
for all $\mathbf{x}$ and
\begin{equation}
\underline{D}(f(\mathbf{X})) \leq \min(\underline{D}(\mathbf{X}),T).
\end{equation}
\end{IEEEproof}

From the literature on fractal geometry and a recent generalization of the coarea formula, we have the following result ensuring the existence of point-wise dimension under a smooth mapping.

\begin{lemma}\label{projection_pointwise}
Let $\mu$ be an absolutely continuous probability measure in $\mathbb{R}^S$. For any smooth function $f: \mathbb{R}^S \to \mathbb{R}^T$, $d(f(\mathbf{x}))$ exists for almost every $f(\mathbf{x})$ on the support of $f(\mu)$.
\end{lemma}

\begin{IEEEproof}
The proof starts by showing that the push forward measure $f(\mu)$ is absolutely continuous with respect to a Hausdorff measure.
From \cite{negro2021}, if $\mu$ is absolutely continuous with respect to the Lebesgue measure, then $f(\mu)$ has a density function for almost every $f(\mathbf{x})$ with respect to the Hausdorff measure $\mathcal{H}^R$, where $R$ is the rank of the Jacobian matrix for the smooth function $f$. This is based on an extension of the coarea formula, which is a generalization of the change of variable formula, to the case of rank deficient mappings in \cite{HAJLASZ2017}. Then applying similar arguments as in \cite[p.21]{cutler1993}, absolutely continuity with respect to $\mathcal{H}^R$ ensures that point-wise dimension exists and has value $R$ for almost every $f(\mathbf{x})$ on the support of $f(\mu)$.
\end{IEEEproof}

The next result is a modification of Corollary 4.2 in \cite{Hunt1997}. It gives
a condition under which the information dimension is preserved under
projection. Specifically, when the point-wise dimension is upper bounded by the
dimension of the range space of the projection, the information dimension is
preserved under projection.
Corollary 4.2 in \cite{Hunt1997} requires the projection to have compact support.
We apply similar techniques as in Lemma \ref{projection_ub} to remove the
requirement for compact support in \cite{Hunt1997}, and instead use the Lipschitz
continuity of linear mappings to arrive at the following:

\begin{lemma}\label{projection}
Given a matrix $\mathbf{A} \in \mathbb{R}^{T\times S}$ and $\mathbf{X}$ with probability measure $\mu\in \mathbb{R}^S$. If the point-wise dimension of $\mu$ exists for almost every $\mathbf{x}$ and $d(\mathbf{x})\leq T$, then for almost all $\mathbf A$ the information dimension of $\mathbf{AX}$ exists and
\begin{equation}
D(\mathbf{A}\mathbf{X})=D(\mathbf{X}).
\end{equation}
\end{lemma}

%

\subsection{Information Dimension of $\mathbf{W}$ via a Connection to the MIMO Phase Noise Channel}

We are now ready to compute the information dimension of the output of the RIS $\mathbf{W}$ in (\ref{systemmodel}). First, consider the following channel model (with general arbitrary dimensions):
\begin{equation}
\mathbf{\tilde Y}= \sqrt{P}{\boldsymbol {\tilde \Theta}} \mathbf{\tilde G} \mathbf{\tilde X}  + \mathbf{\tilde Z},
\end{equation}
where $\mathbf{\tilde Y} \in \mathbb{C}^{\tilde N}$, $\mathbf{\tilde X}\in \mathbb{C}^{\tilde M}$,
${\boldsymbol{\tilde{\Theta}}}={\text{diag}}([e^{j\theta_1},e^{j\theta_2},\cdots,e^{j\theta_{\tilde N}}])$.
A crucial observation is that this channel model resembles the MIMO channel with phase noise. We leverage the following result from the phase noise literature for a key insight.


\begin{lemma}[\cite{Yang2017a}]\label{lemma2}
Let $\boldsymbol{\tilde \Phi}=[e^{j\theta_1},\ldots,e^{j\theta_{\tilde N}}]^{T}$ be a vector of random phases such that $h(\boldsymbol{\tilde \Phi})>- \infty$ and
\begin{equation}
\bar{\mathbf{Y}} = \sqrt{P} {\boldsymbol {\tilde \Phi}} \circ \left(\mathbf{\tilde G} \mathbf{\tilde X}\right)  + \mathbf{\tilde Z},
\end{equation}
where $\circ$ denotes element-wise product and $\mathbf{\tilde Z} \sim\mathcal{CN}(0,\mathbf{I}_{\tilde N})$.

For $\mathbf{\tilde X}\sim \mathcal{CN}(0,\mathbf{I}_{\tilde M})$ independent of $\boldsymbol{\tilde \Phi}$, we have that
for almost all $\mathbf{\tilde G} \in \mathbb{C}^{\tilde N\times \tilde M}$,
\begin{equation}
h(\bar{\mathbf{Y}})\geq \min \left(\tilde{M} + \frac{\tilde{N}}{2} - \frac{1}{2}, \tilde{N} \right)\log^{+}P+ c_1.
\label{eq:h_Y_lowerbound}
\end{equation}
Here $c_1$ is a constant that does not depend on $P$.
\end{lemma}

The intuition behind (\ref{eq:h_Y_lowerbound}) is the following. The information dimension
of the entropy term $h(\bar{\mathbf Y})$ is clearly lower bounded by the dimension of $\mathbf{\bar Y}$
which is $\tilde N$, but it is also lower bounded by the transmit dimension, which is the sum of $\tilde M$ and $\tilde N/2$ but
subtracting $1/2$. Here, $\tilde M$ is the information dimension of $\mathbf{\tilde X}$;
$\tilde N/2$ is the information dimension of the phases $\mathbf{\tilde{\Phi}}$;
but, because there is a common phase between them as the two are multiplied, we need to subtract
the dimension of their overlap, which is $1/2$.

We are now ready to characterize the information dimension of $\mathbf{W}$ in (\ref{systemmodel}).
The result below is for arbitrary dimensions $\tilde N$ and $\tilde M$.
\begin{lemma}\label{theoremN=K}
Consider the channel model
\begin{equation}
\mathbf{\tilde Y}=\sqrt{P}\mathbf{\tilde W}  + \mathbf{\tilde Z} = \sqrt{P}{\boldsymbol {\tilde \Theta}} \mathbf{\tilde G} \mathbf{\tilde X}  + \mathbf{\tilde Z}, \label{systemmodel2}
\end{equation}
where $\mathbf{\tilde Y} \in \mathbb{C}^{\tilde N}$ is the output, $\mathbf{\tilde X}\in \mathbb{C}^{\tilde M}$ with power constraint $\mathbb{E}[\|\mathbf{\tilde X}\|_2^2]\leq 1$ and
${\boldsymbol{\tilde{\Theta}}}={\text{diag}}([e^{j\theta_1},e^{j\theta_2},\cdots,e^{j\theta_{\tilde{N}}}])$ are the inputs, and
 $\mathbf{\tilde Z} \sim\mathcal{CN}(0,\mathbf{I}_{\tilde N})$ is the noise.
For almost all matrices $\mathbf{\tilde G} \in \mathbb{C}^{\tilde N\times \tilde M}$,
the DoF of $I(\mathbf{\tilde X},\boldsymbol{\tilde \Theta};\mathbf{\tilde Y})$ is given by
$\min(\tilde M + \frac{\tilde N}{2}-\frac{1}{2}, \tilde N)$.
This DoF can be achieved using any of the following choices of distributions for $\mathbf{\tilde X}$ and $\boldsymbol{\tilde \Theta}$ with independent $\mathbf{\tilde X}$ and $\boldsymbol{\tilde \Theta}$:
\begin{enumerate}
\item $\mathbf{\tilde X} \sim \mathcal{CN}(0, \mathbf{I}_{\tilde M})$ and $\boldsymbol{\tilde \Theta}$ with i.i.d. $\theta_i$'s and $\theta_i \sim \operatorname{Unif}(-\pi, \pi]$
\item The distribution of $\mathbf{\tilde X}$ is chosen such that the first $\tilde M-1$ elements are i.i.d.\ complex Gaussian, i.e., $\mathbf{\tilde X}_{\tilde M-1}\sim \mathcal{CN}(0,\mathbf{I}_{\tilde M-1})$,
and $\tilde X_{\tilde M}$ is chosen to be real with a chi-squared distribution with 2 degrees of freedom, and $\boldsymbol{\tilde \Theta}$ remains as i.i.d. uniform with  $\tilde \theta_i \sim \operatorname{Unif}(-\pi, \pi]$.
\item The distribution of $\mathbf{\tilde X}$ is i.i.d. complex Gaussian for all $\tilde M$ elements, i.e., $\mathbf{\tilde X}\sim \mathcal{CN}(0,\mathbf{I}_{\tilde M})$,
and the first $\tilde{N}-1$ elements of $\boldsymbol{\tilde \Theta}$ are i.i.d.\ with $\tilde \theta_i \sim \operatorname{Unif}(-\pi, \pi], ~i=1,\ldots, N-1$, and the last element of $\boldsymbol{\tilde \Theta}$ is fixed as 1, i.e., $\tilde \theta_N=1$.
\end{enumerate}
With any of these three choices of distributions, the information dimension of $\mathbf{\tilde W}$ is
\begin{equation}
D(\mathbf{\tilde W})=
\min\left(\tilde M + \frac{\tilde N}{2}-\frac{1}{2}, \tilde N\right).
\end{equation}
\end{lemma}

\begin{IEEEproof}
 We show the achievability of the DoF by making use of Lemma \ref{lemma2}. Observe that with
 $\boldsymbol{\tilde \Phi}=[e^{j\theta_1},\ldots,e^{j\theta_{\tilde N}}]^{T}$,
we have
\begin{align}
\mathbf{\tilde Y} &= \sqrt{P}{\boldsymbol {\tilde \Theta}} \mathbf{\tilde G} \mathbf{\tilde X}  + \mathbf{\tilde Z} \\
&=\sqrt{P} {\boldsymbol {\tilde \Phi}} \circ (\mathbf{\tilde G} \mathbf{\tilde X})  + \mathbf{\tilde Z}.
\end{align}
Therefore, the RIS channel model in \eqref{systemmodel2} can be viewed as a channel with phase shifts
$\boldsymbol{\tilde \Phi}$ applied at the receiver. Choosing $\mathbf{\tilde X} \sim \mathcal{CN}(0, \mathbf{I}_{\tilde M})$ and $\boldsymbol{\tilde \Phi}$ as $\theta_i \sim \operatorname{Unif}(-\pi, \pi]$, we have that $h(\boldsymbol{\tilde \Phi})> -\infty$, therefore we can apply (\ref{eq:h_Y_lowerbound}) in Lemma \ref{lemma2} to get $h(\mathbf{\tilde Y})\geq \min(\tilde M + \frac{\tilde N}{2}-\frac{1}{2}, \tilde N) \log^{+}P+ c_1$.
The achievability of the desired DoF now follows by lower bounding
the mutual information as follows:
\begin{align}
I (\mathbf{\tilde X},\boldsymbol{\tilde \Theta};\mathbf{\tilde Y}) &= h(\mathbf{\tilde Y})-h(\mathbf{\tilde Y}|\mathbf{\tilde X},\boldsymbol{\tilde \Theta})\\
& \geq
\min\left(\tilde M + \frac{\tilde N}{2}-\frac{1}{2}, \tilde N\right) \log^{+}P+ c_1,
\label{eq:I_lower}
\end{align}
where the inequality comes from the fact that
$h(\mathbf{\tilde Y}|\mathbf{\tilde X},\boldsymbol{\tilde \Theta})$ is the entropy of the noise term which does not depend on $P$.

Next, we prove the converse by giving an upper bound on information dimension of $\mathbf{\tilde W}$. Note that we cannot immediately apply Lemma \ref{projection_ub} here as the mapping from $(\mathbf{ \tilde X}, \mathbf{\tilde \Theta})$ to $\mathbf{\tilde W}$ is not Lipschitz continuous. To resolve this, we split the magnitude and phase of $\mathbf{\tilde W}$ and consider them individually.

Denote $|\mathbf{\tilde W}|$ and $\angle \mathbf{\tilde W}$ as the magnitudes and phases of the elements of $\mathbf{\tilde W}$, respectively, where $|\cdot|$ and $\angle \cdot$ are element-wise operations on $\mathbf{\tilde W}$. Then we observe that $|\mathbf{\tilde W}|$ depends only on $\mathbf{\tilde X}$, and the mapping from $\mathbf{\tilde X}$ to $|\mathbf{\tilde W}|$ is equivalent to the mapping from $\mathbf{\tilde X}$ to $|\mathbf{\tilde G\tilde X}|$, which is a Lipschitz continuous mapping and we denote this mapping as $f$.

Then, for any $\mathbf{\tilde X}\in \mathbb{C}^{\tilde M}$, denote the phase of the last element of $\mathbf{\tilde X}$ as $e^{j\theta_{\tilde M}}$. We construct a new $\mathbf{\tilde X}'=e^{-j\theta_{\tilde M}}\mathbf{\tilde X}$ and have $\overline {D}(\mathbf{\tilde X}')\leq \tilde M-\frac{1}{2}$ since the phase of the last element is constant which eliminates one real dimension.
With this construction, since the distribution of $f(\mathbf{\tilde X})$ and $f(\mathbf{\tilde X}')$ are the same, we have 
\begin{align}
\overline {D}(f(\mathbf{\tilde X}))&=\overline {D}(f(\mathbf{\tilde X}')) \nonumber \\
&\leq \min \left(\overline {D}(\mathbf{\tilde X}'), \frac{\tilde N}{2}\right)\leq \min \left(\tilde M-\frac{1}{2}, \frac{\tilde N}{2}\right),
\end{align}
where we apply Lemma \ref{projection_ub} in the first inequality.

On the other hand, $\overline{D} (\angle \mathbf{\tilde W})\leq \frac{\tilde N}{2}$ since it is on $\mathbb{R}^{\tilde N}$.
Putting together, we obtain
\begin{equation}
\overline {D} (\mathbf{\tilde W})\leq \overline {D} (|\mathbf{\tilde W}|)+\overline {D} (\angle \mathbf{\tilde W})\leq \min\left(\tilde M+\frac{\tilde N}{2}-\frac{1}{2}, \tilde N \right).
\end{equation}

In view of Lemma \ref{connection}, this yields the following upper bound on the mutual information
$I (\mathbf{\tilde X},\boldsymbol{\tilde \Theta};\mathbf{\tilde Y})$,
\begin{equation}\label{eq:I_upper}
I (\mathbf{\tilde X},\boldsymbol{\tilde \Theta};\mathbf{\tilde Y})\leq \min\left(\tilde M+\frac{\tilde N}{2}-\frac{1}{2}, \tilde N\right) \log^+ P + c_2.
\end{equation}

Combining (\ref{eq:I_lower}) and (\ref{eq:I_upper}), we see that
the DoF of the channel model \eqref{systemmodel2} must be given by
$\min(\tilde M + \frac{\tilde N}{2}-\frac{1}{2}, \tilde N)$.
The information dimension of $\mathbf{\tilde W}=\boldsymbol{\tilde \Theta}\mathbf{\tilde G\tilde X}$ is therefore $D(\mathbf{\tilde W}) = \min(\tilde M + \frac{\tilde N}{2}-\frac{1}{2}, \tilde N)$,
in view of Lemma \ref{connection}.

The proof so far has used the distribution in Case 1) in the lemma statement, i.e.,
$\mathbf{\tilde X} \sim \mathcal{CN}(0, \mathbf{I}_{\tilde M})$ and $\boldsymbol{\tilde \Theta}$ with i.i.d. $\tilde \theta_i \sim \operatorname{Unif}(-\pi, \pi]$ to achieve the information dimension of $\mathbf{\tilde W}=\boldsymbol{\tilde \Theta}\mathbf{\tilde G \tilde X}$ as 
$\min(\tilde M + \frac{\tilde N}{2}-\frac{1}{2}, \tilde N)$.

Denote the phase of the last element of $\mathbf{\tilde X}$ as $e^{j\phi}$, and the phase of the last element of $\tilde {\boldsymbol{\Theta}}$ as $e^{j\phi'}$.
Here, $\phi$ and $\phi'$ are independent random variables.

For Case 2), we chose $\mathbf{\tilde X'}$ as $e^{-j\phi}\mathbf{\tilde X}$ and $\boldsymbol{\tilde \Theta'}=e^{j\phi}\boldsymbol{\tilde \Theta}$. We have that the distribution of $\boldsymbol{\tilde \Theta'}\mathbf{\tilde G\tilde X'}$ is the same as the distribution of $\boldsymbol{\tilde \Theta}\mathbf{\tilde G \tilde X}$, so the same DoF and also the same information dimension for $\mathbf{\tilde W'}=\boldsymbol{\tilde \Theta'}\mathbf{\tilde G\tilde X'}$ can be achieved. For this choice of distributions, $\mathbf{\tilde X}'$ is such that the first $\tilde M-1$ elements are i.i.d. complex Gaussian, i.e. $\mathbf{\tilde X}'_{\tilde M-1}\sim \mathcal{CN}(0,\mathbf{I}_{\tilde M-1})$, the real part of $X'_{\tilde M}$ is a chi-squared distribution with 2 degrees of freedom, and the imaginary part of $X'_{\tilde M}$ is $0$, and $\boldsymbol{\tilde \Theta'}$ as i.i.d.\ $\tilde{\theta}'_i \sim \operatorname{Unif}(-\pi, \pi]$.
Note that since the phase of the last component of $\mathbf{\tilde X'}$ is deterministic, $\mathbf{\tilde X'}$ and $\boldsymbol{\tilde \Theta'}$ remain independent.

For Case 3), we chose $\mathbf{\tilde X'}$ as $e^{j\phi'}\mathbf{\tilde X}$ and $\boldsymbol{\tilde \Theta'}=e^{-j\phi'}\boldsymbol{\tilde \Theta}$. Similarly, the distribution of $\boldsymbol{\tilde \Theta'}\mathbf{\tilde G\tilde X'}$ is the same as the distribution of $\boldsymbol{\tilde \Theta}\mathbf{\tilde G \tilde X}$, so the same DoF and the same information dimension for $\mathbf{\tilde W'}=\boldsymbol{\tilde \Theta'}\mathbf{\tilde G\tilde X'}$ can be achieved. For this choice of distributions, $\mathbf{\tilde X}'$ is still a standard complex Gaussian, i.e. $\mathbf{\tilde X}'_{\tilde M}\sim \mathcal{CN}(0,\mathbf{I}_{\tilde M})$ since multiplying by a phase does not change the distribution of a circularly symmetric complex Gaussian vector. The distribution of the first $\tilde N -1$ elements of $\boldsymbol{\tilde \Theta'}$ remains unchanged as $\tilde{\theta}'_i \sim \operatorname{Unif}(-\pi, \pi]$, because $e^{j\phi'}$ is a periodic function with period $2\pi$, so the distribution is unchanged after a shift of $2\pi$.
Note again that $\mathbf{\tilde X'}$ and $\boldsymbol{\tilde \Theta'}$ remain independent.

As a result, all three distributions achieve the same DoF and the same information dimension.
\end{IEEEproof}

\subsection{Proof of Theorem \ref{theoremNneqK}}

Now we are ready to prove Theorem \ref{theoremNneqK}. First, we establish that (\ref{eq:HW_dimension}) is an upper bound for $D(\mathbf{HW})$. From Lemma \ref{theoremN=K} and Lemma \ref{connection} we have that $D(\mathbf{W}) \leq \min(M+\frac{N}{2}-\frac{1}{2}, N)$ for any distribution of $\boldsymbol{\Theta}$ and $\mathbf{X}$. Then, by Lemma \ref{projection_ub}, we have $D(\mathbf{HW})\leq \min(M+ \frac{N}{2}-\frac{1}{2}, N, K)$ since linear transformations are Lipschitz continuous.

To establish the lower bound, i.e., $D(\mathbf{HW})\geq \min(M+ \frac{N}{2}-\frac{1}{2}, N, K)$,
we divide into two cases and first treat the case of $K \geq \min(M+\frac{N}{2}-\frac{1}{2}, N, K)$.
The information dimension of $\mathbf{W}$ is $D(\mathbf{W}) = \min(M+ \frac{N}{2}-\frac{1}{2}, N)$ in view of Lemma \ref{theoremN=K}.
The key idea is to recognize that because $\mathbf{X}$ and $\mathbf{\Theta}$ have
$\frac{1}{2}$ dimension overlap, we should choose one of the two distributions for $(\mathbf{X},\boldsymbol{\Theta})$
given in Lemma \ref{theoremN=K} that explicitly account for the overlap, i.e., the distributions in either Case 2) or 3).
In either case, $(\mathbf{X}, \boldsymbol{\Theta})$ has an absolutely continuous distribution of dimension $M+\frac{N}{2}-\frac{1}{2}$, so $(\mathbf{X}, \boldsymbol{\Theta})$ has a point-wise dimension
$d(\mathbf{X}, \boldsymbol{\Theta})=M+\frac{N}{2}-\frac{1}{2}$ for every point.
Now, denote the mapping from $(\mathbf{X}, \boldsymbol{\Theta})$ to $\mathbf{W}$ as $f$,
which is smooth. Using the same technique of decomposing $\mathbf{W}$ to $|\mathbf{W}|$ and $\angle \mathbf{W}$ as in the proof of Lemma \ref{theoremN=K} and applying Lemma \ref{projection_ub}, we have the following upper bound on the point-wise dimension:
\begin{align}
\overline{d}(\mathbf{W})&=\overline{d}(f(\mathbf{X},\boldsymbol{\Theta})) \\
&\leq \min \left(\overline{d}(\mathbf{X},\boldsymbol{\Theta}),N\right)=\min \left(M+\frac{N}{2}-\frac{1}{2}, N\right).
\end{align}
Thus, for all
$K \geq \min(M+\frac{N}{2}-\frac{1}{2}, N)$, we can now apply Lemma \ref{projection_pointwise} and Lemma \ref{projection} to obtain
\begin{align}
D(\mathbf{HW})=D(\mathbf{W})=\min\left(M+\frac{N}{2}-\frac{1}{2}, N\right).
\end{align}

For the case of $K < \min(M+\frac{N}{2}-\frac{1}{2}, N)$, the idea is to choose $\tilde M \leq M$ and $\tilde N\leq N$ such that $K = \min(\tilde M + \frac{\tilde N}{2}-\frac{1}{2}, \tilde N)$ and use such $(\tilde M, \tilde N)$ in Lemma \ref{theoremN=K}.
In effect, we are choosing a distribution for $(\mathbf{X, \Theta})$ by
turning off some of the transmit antennas and set some of the reflecting elements to be deterministic.

First we provide a characterization of the information dimension of 
$\mathbf{W}'=\boldsymbol {\Theta} \mathbf{G} \mathbf{X}$ where  $\boldsymbol{\Theta}=\left[ \begin{array}{cc}
	{\boldsymbol {\Theta}'} & \mathbf{0} \\
	\mathbf{0} & \mathbf{I}_r \\
\end{array} \right]$,
this corresponds to the scenario where $r$ out of $N$ reflective elements are set to be deterministic. For the case of $r=1$, we have already established that $D(\mathbf{W'}) = \min(M+\frac{N-1}{2}, N)$ (see Case 3 in Lemma 6). Now define
$\mathbf{Y'}
= \sqrt{P} {\boldsymbol {\Theta}} \mathbf{G} \mathbf{X} + \mathbf{Z}
= \sqrt{P} \mathbf{W'} + \mathbf{Z}$.
Since
\begin{equation}
	I(\mathbf{X}, \boldsymbol{\Theta}; \mathbf{Y'})
	= I(\mathbf{X}; \mathbf{Y'}) + I(\boldsymbol{\Theta}; \mathbf{Y'}|\mathbf{X}),
	\label{eq:chain_rule_theta1}
\end{equation}
and the DoF corresponding to the $I(\mathbf{\Theta};\mathbf{Y'}|\mathbf{X})$
term is $\frac{N-1}{2}$, we have that the DoF corresponding to
the $I(\mathbf{X};\mathbf{Y'})$ term is $\min(M,\frac{N+1}{2})$.

To generalize the above results to the $r>1$ case, we proceed to show that
the DoF corresponding to $I(\mathbf{X};\mathbf{Y'})$ is
$\min(M,\frac{N}{2}+\frac{r}{2})$ for $N\geq r>1$.
Toward this end, we define $\mathbf{Y'} = [\mathbf{Y}_1 \mathbf{Y}_2]$, where
$\mathbf{Y}_1$ corresponds to the first $N-r$ entries, and
$\mathbf{Y}_2$ corresponds to the last $r$ entries, and write
\begin{equation}
	I(\mathbf{X}; \mathbf{Y'})
	= I(\mathbf{X}; \mathbf{Y}_2) + I(\mathbf{X}; \mathbf{Y}_1|\mathbf{Y}_2).
	\label{eq:chain_rule_Y1}
\end{equation}
We design a precoder $\mathbf{X} = \mathbf{Q} \mathbf{X'}$ where
$\mathbf{Q}$ is an unitary $M \times M$ matrix such that $\mathbf{GQ}$ has an upper-triangular
structure. This is possible by using an RQ-decomposition \cite{horn_matrix_analysis2012} on the matrix $\mathbf{G}$. Since $\mathbf{Q}$ is unitary, the resulting $\mathbf{X}$ has the same distribution as $\mathbf{X'}$.

Further, we partition $\mathbf{X'} = [\mathbf{X}'_1 \mathbf{X}'_2]$,
where $\mathbf{X}'_1$ corresponds to the first $M-r+1$ entries,
and $\mathbf{X}'_2$ corresponds to the rest of $r-1$
entries.
In this way, $\mathbf{X}'_1$ does not interfere with $\mathbf{X}'_2$, so that
assuming i.i.d.~Gaussian distribution on $\mathbf{X'}$,
a DoF of $r$ corresponding to the first term of
(\ref{eq:chain_rule_Y1}) can be achieved with
$I(\mathbf{X}'_2;\mathbf{Y}_2)$, because this is a usual MIMO channel in which both input dimension and output dimension are $r-1$.

For the second term in (\ref{eq:chain_rule_Y1}),
since $\mathbf{X}'_2$ is decoded at the receiver based on $\mathbf{Y}_2$,
its effect can be subtracted from $\mathbf{Y}_1$; 
so the second term in (\ref{eq:chain_rule_Y1}) is reduced to the case of $r=1$, where the input is now $\mathbf{X}'_1$
of dimension $M-r+1$ and output is $\mathbf{Y}_1$ of dimension $N-r+1$. Using the result from the previous $r=1$ case,
a DoF of $\min(M-r+1, \frac{N-r+1+1}{2})$ can be achieved.
Putting the two terms together, we achieve a DoF of
$\min(M,\frac{N+r}{2})$ for $I(\mathbf{X}; \mathbf{Y}')$.

Thus overall, a DoF of $\min(M, \frac{N}{2}+r)$ is achievable for
$I(\mathbf{X}, \mathbf{Y'})$ for all $r \geq 1$..
Then, by the chain rule (\ref{eq:chain_rule_theta1}), we have that
the DoF of $I(\mathbf{X}, \boldsymbol{\Theta'}; \mathbf{Y'})$ and also
the information dimension $D(\mathbf{W}')$ are at least
$\min\left(M+\frac{N-r}{2},N\right)$.

On the other hand, because $\mathbf{W'}$
is a mapping from
$(\mathbf{X},\boldsymbol{\Theta})$, the information dimension
and the point-wise dimension of $\mathbf{W'}$ are at most
$M+\frac{N-r}{2}$. Additionally,
the information dimension is also upper bounded by the dimension of
$\boldsymbol {\Theta}$, which is $N$.
Therefore, $D(\mathbf{W}')$ is at most $\min\left(M+\frac{N-r}{2},N\right)$.

This establishes that by setting $r\geq 1$ reflecting elements to be deterministic, the information dimension of the overall input $\mathbf{W}'$ is
$D(\mathbf{W}') = \min\left(M+\frac{N-r}{2},N\right)$. Therefore we can choose an appropriate $N \geq r \geq 0$ and $\tilde M \leq M$ such that $\min\left(\tilde M+\frac{N-r}{2},N\right)=K$.
Then we apply the same argument as in the case of $K \geq \min(M+\frac{N}{2}-\frac{1}{2}, N)$, which gives 
\begin{align}
	\overline{d}(\mathbf{W})&=\overline{d}(f(\mathbf{X}, \boldsymbol{\Theta})) \\
	&\leq \min \left(\overline{d}(\mathbf{X}, \boldsymbol{\Theta}),\tilde N\right)=\min \left(\tilde M + \frac{\tilde N}{2}-\frac{1}{2}, \tilde N\right).
\end{align}
Then, the proof follows by applying Lemma \ref{projection_pointwise} and Lemma \ref{projection} to obtain that $D(\mathbf{HW})=D(\mathbf{W})=\min(\tilde M+\frac{\tilde N}{2}-\frac{1}{2}, \tilde N)=K$ is achievable.

Combining the two cases gives us the achievability. Together with the converse and by making use of Lemma \ref{connection}, we obtain that the DoF of \eqref{systemmodel} is given by
\begin{equation}
\mathrm{DoF}_{(\mathbf X,\boldsymbol{\Theta})} = D(\mathbf{HW})=\min \left(M+ \frac{N}{2}-\frac{1}{2}, N, K \right),
\end{equation}
thus proving Theorem \ref{theoremNneqK}.

\section{Proof of Theorem \ref{theorem_knownX}}\label{app:proof_DoFknownx}

The proof follows the same line as the proof of Theorem \ref{theoremNneqK}.
First, from the fact that $\boldsymbol{\Phi}$ is constrained to have constant amplitude, its information dimension is upper bounded by
\begin{equation}
D(\boldsymbol{\Phi})\leq \frac{N}{2},
\end{equation}
with equality achieved by choosing absolutely continuous distributions for $\boldsymbol{\Phi}$. From Lemma \ref{projection_ub}, we obtain an upper bound for the information dimension of $\bar{\mathbf{H}} \boldsymbol{\Phi}$ as
\begin{equation}
D(\bar{\mathbf{H}} \boldsymbol{\Phi}) \leq \min \left(\frac{N}{2},K\right).
\end{equation}
Next we show that this upper bound can be achieved. To do this, choose $\boldsymbol{\Phi}$ to have an absolutely continuous distributions in each of its elements, e.g., $\operatorname{Unif}(-\pi,\pi]$. With this choice of $\boldsymbol{\Phi}$, not only the information dimension of $\boldsymbol{\Phi}$ is $\frac{N}{2}$, the point-wise dimension of $\boldsymbol{\Phi}$ is also $\frac{N}{2}$ for every point.

Consider two cases. If $K \geq \frac{N}{2}$, we can directly apply Lemma \ref{projection} to obtain
\begin{align}
D(\bar{\mathbf{H}} \boldsymbol{\Phi})=D(\boldsymbol{\Phi})=\frac{N}{2}.
\end{align}

If $K <\frac{N}{2}$, we choose $\tilde {N} \leq N$ such that $K = \frac{\tilde{N}}{2}$ and obtain a new $\tilde {\boldsymbol{\Phi}}$ with $d(\tilde{\boldsymbol{\Phi}})=\frac{\tilde{N}}{2}$.
In effect, we set some of the reflecting elements to
have deterministic phases.
Then we apply the same argument as in the case of $K \geq \frac{N}{2}$ to get
\begin{align}
d(\bar{\mathbf{H}}\tilde{\boldsymbol{\Phi}})=\frac{N}{2}=K.
\end{align}
Then the proof follows by applying Lemmas \ref{projection} to obtain that $D(\bar{\mathbf{H}}\tilde{\boldsymbol{\Phi}})=\frac{N}{2}=K$ is achievable.

Combining the two cases gives us the achievability. Together with the converse and by making use of Lemma \ref{connection}, we complete the proof.

\section{Proof of the General Case in Theorem \ref{theorem_direct_path}} \label{app:proof_DoFdirect}
In this appendix, we prove Theorem \ref{theorem_direct_path} for the general case of $r>1$. Recall that the channel model
\begin{align}
\mathbf{Y}
= \sqrt{P}\mathbf{H'} \underbrace{{\boldsymbol {\Theta}'} \mathbf{G'} \mathbf{X}}_{\mathbf{W'}} + \mathbf{Z},
\end{align}
with $\boldsymbol{\Theta'}=\left[ \begin{array}{cc}
	{\boldsymbol {\Theta}} & \mathbf{0} \\
	\mathbf{0} & \mathbf{I}_r \\
\end{array} \right]$ is effectively that of a system with RIS of size
$N'=N+r$ with $r$ elements fixed.
The idea is to follow the proof of Theorem \ref{theoremNneqK}. We first characterize the information dimension of $D(\mathbf{W'})$, then study its behavior under projection.
Throughout the proof, we assume that $\mathbf{X}$ and $\boldsymbol{\Theta}$ are independent.

For the case of $r=1$, we have already established that the DoF of
joint transmission of $(\mathbf{X}, \boldsymbol{\Theta})$ to $\mathbf Y$
is $\min(M+\frac{N}{2}, N+1)$.
This means that $D(\mathbf{W'}) = \min(M+\frac{N}{2}, N+1)$.
Define
$\mathbf{Y'}
= \sqrt{P} {\boldsymbol {\Theta}'} \mathbf{G'} \mathbf{X} + \mathbf{Z}
= \sqrt{P} \mathbf{W'} + \mathbf{Z}$.
Since
\begin{equation}
I(\mathbf{X}, \boldsymbol{\Theta'}; \mathbf{Y'})
= I(\mathbf{X}; \mathbf{Y'}) + I(\boldsymbol{\Theta'}; \mathbf{Y'}|\mathbf{X}),
\label{eq:chain_rule_theta}
\end{equation}
and the DoF corresponding to the $I(\mathbf{\Theta}';\mathbf{Y}'|\mathbf{X})$
term is $\frac{N}{2}$, we have that the DoF corresponding to
the $I(\mathbf{X};\mathbf{Y'})$ term is $\min(M,\frac{N}{2}+1)$.

To generalize the above results to the $r>1$ case, we proceed to show that
the DoF corresponding to $I(\mathbf{X};\mathbf{Y'})$ is
$\min(M,\frac{N}{2}+r)$ for $r>1$.
Toward this end, we define $\mathbf{Y'} = [\mathbf{Y}_1 \mathbf{Y}_2]$, where
$\mathbf{Y}_1$ corresponds to the first $N+1$ entries, and
$\mathbf{Y}_2$ corresponds to the last $r-1$ entries, and write
\begin{equation}
I(\mathbf{X}; \mathbf{Y'})
	= I(\mathbf{X}; \mathbf{Y}_2) + I(\mathbf{X}; \mathbf{Y}_1|\mathbf{Y}_2).
\label{eq:chain_rule_Y'}
\end{equation}
We design a precoder $\mathbf{X} = \mathbf{Q} \mathbf{X'}$ where
$\mathbf{Q}$ is a unitary $M \times M$ matrix such that $\mathbf{G' Q}$ has an upper-triangular
structure. This is possible by using an RQ-decomposition \cite{horn_matrix_analysis2012} on the matrix $\mathbf{G'}$. Since $\mathbf{Q}$ is unitary, the resulting $\mathbf{X}$ has the same distribution as $\mathbf{X'}$.

Further, we partition $\mathbf{X'} = [\mathbf{X}'_1 \mathbf{X}'_2]$,
where $\mathbf{X}'_1$ corresponds to the first $M-r+1$ entries,
and $\mathbf{X}'_2$ corresponds to the rest of $r-1$
entries.
In this way, $\mathbf{X}'_1$ does not interfere with $\mathbf{X}'_2$, so that
assuming i.i.d.~Gaussian distribution on $\mathbf{X'}$,
a DoF of $r-1$ corresponding to the first term of
(\ref{eq:chain_rule_Y'}) can be achieved with
$I(\mathbf{X}'_2;\mathbf{Y}_2)$, because this is a usual MIMO channel in which both input dimension and output dimension are $r-1$.

For the second term in (\ref{eq:chain_rule_Y'}),
since $\mathbf{X}'_2$ is decoded at the receiver based on $\mathbf{Y}_2$,
its effect can be subtracted from $\mathbf{Y}_1$; 
so the second term in (\ref{eq:chain_rule_Y'}) is reduced to the case of $r=1$, where the input is now $\mathbf{X}'_1$
of dimension $M-r+1$ and output is $\mathbf{Y}_1$ of dimension $N+1$. Using the result from the previous $r=1$ case,
a DoF of $\min(M-r+1, \frac{N}{2}+1)$ can be achieved.
Putting the two terms together, we achieve a DoF of
$\min(M,\frac{N}{2}+r)$ for $I(\mathbf{X}; \mathbf{Y}')$.

Thus overall, a DoF of $\min(M, \frac{N}{2}+r)$ is achievable for
$I(\mathbf{X}, \mathbf{Y'})$ in the $r>1$ case.
Then, by the chain rule (\ref{eq:chain_rule_theta}), we have that
the DoF of $I(\mathbf{X}, \boldsymbol{\Theta'}; \mathbf{Y'})$ and also
the information dimension $D(\mathbf{W}')$ are at least
$\min\left(M+\frac{N}{2},N+r\right)$.

On the other hand, because $\mathbf{W'}$
is a mapping from
$(\mathbf{X},\boldsymbol{\Theta})$, the information dimension
and the point-wise dimension of $\mathbf{W'}$ are at most
$M+\frac{N}{2}$. Additionally,
the information dimension is also upper bounded by the dimension of
$\boldsymbol {\Theta}'$, which is $N+r$.
Therefore, $D(\mathbf{W}')$ is at most $\min\left(M+\frac{N}{2},N+r\right)$.

This establishes that
$D(\mathbf{W}') = \min\left(M+\frac{N}{2},N+r\right)$.
Finally, by the same argument as in the proof of Theorem \ref{theoremNneqK},
we can show that after projection under $\mathbf{H}'$, the DoF of
the overall channel is $\min\left( M+\frac{N}{2}, N+r, K \right)$.

\section{Augmented Lagrangian with Riemannian Conjugate Gradient Method for Solving (\ref{slp2})}
\label{app:riemannian}

In this appendix, we provide the implementation details of the proposed
algorithm for solving (\ref{slp2}). For the ease of discussion, we use
the vector form of
$\mathbf{\Theta}$, i.e.,
$\boldsymbol{\Phi}=[e^{j\theta_1},\ldots,e^{j\theta_{N}}]^{T}$.
Then the objective function of (\ref{slp2}) can be written as
\begin{equation}
f(\mathbf{X},\boldsymbol{\Phi})={\|\hat{\mathbf{Y}}-\sqrt{P}\left(\mathbf{H}\mathrm{diag}(\boldsymbol {\Phi})\mathbf{G}+\mathbf{F}\right)\mathbf{X}\|_2^2}.
\end{equation}
The constraints are
the unit modulus constraints on the element of $\boldsymbol{\Phi}$,
and the power constraint on $\mathbf{X}$, i.e., $\|\mathbf{X}\|_2^2\leq 1$.

The main steps of the algorithm follow the augmented Lagrangian approach
\cite{alm_boumal2020} for accounting for the power constraint, and
the Riemannian conjugate gradient method \cite{manifold_IRS2019} for accounting
for the unit modulus constraints.

\subsection{Augmented Lagrangian Method}

We adopt the augmented Lagrangian approach, which has a convergence guarantee,
as a method to move the constraint to the objective function.
The augmented Lagrangian function is defined as
\begin{equation} \label{aug_lagrangian}
L(\mathbf{X},\boldsymbol{\Phi},\lambda,\rho)=f(\mathbf{X},\boldsymbol{\Phi})+\frac{\rho}{2}\max\left(0, \frac{\lambda}{\rho}+\|\mathbf{X}\|_2^2-1\right)^2
\end{equation}
The augmented Lagrangian method alternates between updating the primal variables $(\mathbf{X},\boldsymbol{\Phi})$ and the dual variable $\lambda$. For updating of $(\mathbf{X},\boldsymbol{\Phi})$, we adopt the conjugate gradient method on the Riemannian manifold to minimize
\eqref{aug_lagrangian} for a fixed $\lambda$, the obtained solution is then used as the initial point for the next update. For updating $\lambda$, we adopt the clipped gradient type update rule \cite{alm_2014}. The overall procedure for solving \eqref{slp2} is provided in Algorithm \ref{algo:alm}.

In Algorithm \ref{algo:alm}, the clip function is defined as follows
\begin{equation}
\operatorname{clip}_{[\alpha,\beta]}(x)=\max(\alpha, \min(\beta, x)).
\end{equation}
There are also a number of constants that have to be chosen for the algorithm
to run smoothly. The initial accuracy $\epsilon_0$ and the accuracy tolerance
$\epsilon_{\min}$ control the threshold on the norm of the gradient under which
we consider \eqref{alm_sub} to be solved. The accuracy controlling constant
$\theta_\epsilon \in (0,1)$ decreases the accuracy threshold in every iteration.
The initial penalty coefficient $\rho_0$ controls the level of penalization for
violating the constraints, and it is controlled by the penalty increasing
factor $\theta_\rho > 1$. The parameter $\theta_{\sigma}\in (0,1)$ controls
when the penalty coefficient $\rho$ needs to be updated; it ensures that $\rho$
is updated only when the amount of constraint violation is shrinking fast
enough. Finally, the minimum step size $d_{\min}$ determines the minimum
change between two iterates for the algorithm to continue. Here we use
the Euclidean norm to measure the amount of change between the iterates, i.e.
$\mathrm{dist}(x,y)=\|x-y\|_2^2$. The parameters used for
implementing Algorithm \ref{algo:alm} are: $\epsilon_0=10^{-3}$,
$\epsilon_{\min}=10^{-6}$, $\theta_\epsilon=1000^{-\frac{1}{30}}$, $\rho_0=1$,
$\theta_\rho =10 $, $\theta_{\sigma}=0.8$, $\lambda_0=1$,
$\lambda^{\max}=10000$, and $d_{\min}=10^{-6}$.

Algorithm \ref{algo:alm} is guaranteed to converge to a stationary problem of the original problem \eqref{slp2} under mild conditions. For the convergence analysis of the augmented Lagrangian method on manifolds, we refer the reader to \cite{alm_2014} and \cite{alm_boumal2020}.

\begin{algorithm}[t]\label{algo:alm}
\SetAlgoLined
\SetKw{KwInput}{Input:}
\KwInput{
\begin{itemize}
\item Starting point $(\mathbf{X}_0,\boldsymbol{\Phi}_0)$
\item Initial value for the Lagrangian multiplier $\lambda_0$
\item Accuracy tolerance $\epsilon_{\min}$
\item Initial accuracy $\epsilon_0$
\item Initial penalty coefficient $\rho_0$
\item Accuracy controlling constant $\theta_\epsilon \in (0,1)$
\item Penalty increasing factor $\theta_\rho > 1$
\item Multiplier boundaries $\lambda^{\max}$
\item Constraint violation control $\theta_{\sigma}\in (0,1)$
\item Minimum step size $d_{\min}$
\end{itemize}
}
 \For{$k = 0,1,\dots$}
 {
    Compute $(\mathbf{X}_{k+1}, \boldsymbol{\Phi}_{k+1})$, an approximate solution to the following problem within a tolerance $\epsilon_{k}$:
    \begin{equation}\label{alm_sub}
    \begin{aligned}
    & \underset{\mathbf{X}, \boldsymbol{\Phi}\in {\mathbb{M}_{cc}^N}}{\text{min}}
    & & L(\mathbf{X},\boldsymbol{\Phi}, \lambda_k,\rho_k)\\
    \end{aligned}
    \end{equation}
    with $(\mathbf{X}_k,\boldsymbol{\Phi}_k)$ as the starting point.

    \If{$\mathrm{dist}((\mathbf{X}_k,\boldsymbol{\Phi}_k), (\mathbf{X}_{k+1},\boldsymbol{\Phi}_{k+1})) < d_{\min}$ $\mathrm{and}$ $\epsilon_k \leq \epsilon_{\min}$}{
        Return $(\mathbf{X}_{k+1},\boldsymbol{\Phi}_{k+1})$\;
    }
    $\lambda_{k+1}= \operatorname{clip}_{[0,\lambda^{\max}]}(\lambda_k + \rho_k (\|\mathbf{X}_{k+1}\|_2^2-1))$ \;
    $\sigma_{k+1}= \max\left\{ \left(\|\mathbf{X}_{k+1}\|_2^2-1\right), - \frac{\lambda_{k}}{\rho_k}\right\}$ \;
    $\epsilon_{k+1} = \max\left\{\epsilon_{\min}, \theta_\epsilon \epsilon_k\right\}$\;
    \eIf{$k=0$ $\mathrm{or}$ $|\sigma_{k+1}| \leq \theta_{\sigma}|\sigma_{k}|$}
    {
        $\rho_{k+1} = \rho_k$\;
    }
    {
    	$\rho_{k+1} = \theta_\rho \rho_k$\;
	}
 }
\caption{Augmented Lagrangian Method for Solving \eqref{slp2}}
\end{algorithm}

\subsection{Riemannian Conjugate Gradient}

For solving the subproblem \eqref{aug_lagrangian}, the conjugate gradient approach on the manifold is adopted. For the implementation of the Riemannian conjugate gradient, the Matlab package \emph{Manopt}\cite{manopt} is used. This appendix describes the basic principles of the algorithm.
The constraint on each element of $\boldsymbol{\Phi}$ can be regarded as the complex unit circle manifold as
\begin{equation}
\mathbb{M}_{cc}=\{\theta_i\in\mathbb{C}:\theta_i^*\theta_i=1\}.
\end{equation}
For a given point $\theta_i$ on the manifold $\mathbb{M}_{cc}$, the directions along which it can move are characterized by its tangent space. The tangent space at point $\theta_i \in\mathbb{M}_{cc}$ is represented by
\begin{equation}
T_{\theta_i}\mathbb{M}_{cc}=\{z\in\mathbb{C}:z^*\theta_i+\theta_i^*z=2\left\langle {\theta_i,z} \right\rangle=0\}.
\end{equation}

In \eqref{aug_lagrangian}, the vector $\boldsymbol{\Phi}$ is constrained to be in a
product manifold of $N$ complex circle manifolds, i.e.,
$\mathbb{M}_{cc}^N=\{{\boldsymbol{\Phi}\in\mathbb{C}^N:|\theta_1|=|\theta_2|=\ldots=|\theta_N|=1}\}$,
which is a Riemannian submanifold of $\mathbb{C}^N$.
With this geometry, the tangent space at a given
point $\boldsymbol{\Phi}\in\mathbb{M}_{cc}^N$ can be expressed as
\begin{equation}
T_{\boldsymbol{\Phi}}\mathbb{M}_{cc}^N=\left\{\mathbf{z}\in\mathbb{C}^N:\Re\left\{\mathbf{z}\circ \boldsymbol{\Phi}^*\right\}=\mathbf{0}_N\right\}.
\end{equation}

Similar to the Euclidean space, the tangent vector along the direction of the negative \emph{Riemannian gradient} represents a direction of steepest descent 
(under the {Riemannian metric} defined on the tangent space, which in this case is just the Euclidean norm while the descent direction is constrained to be within the tangent space).
For the objective function $L(\mathbf{X},\boldsymbol{\Phi},\lambda,\rho)$ in (\ref{aug_lagrangian}), the Riemannian gradient with respect to $\boldsymbol{\Phi}$, denoted as $\mathrm{grad}_{\boldsymbol{\Phi}} L$, is given by the projection of the Euclidean gradient $\nabla_{\boldsymbol{\Phi}} L$ onto the tangent space $T_{\boldsymbol{\Phi}}\mathbb{M}_{cc}^N$:
\begin{equation}\label{rgradient}
\begin{split}
\mathrm{grad}_{\boldsymbol{\Phi}} L&=\mathrm{Proj}_{T_{\boldsymbol{\Phi}}\mathbb{M}_{cc}^N}
\nabla_{\boldsymbol{\Phi}} L \\
&=\nabla_{\boldsymbol{\Phi}} L - \Re\{\nabla_{\boldsymbol{\Phi}} L \circ \boldsymbol{\Phi}^{*}\}\circ\boldsymbol{\Phi},
\end{split}
\end{equation}
where the Euclidean gradient $\nabla_{\boldsymbol{\Phi}} L$ is
\begin{multline}
\nabla_{\boldsymbol{\Phi}} L = -2\sqrt{P}\left(\mathbf{H} \mathrm{diag} (\mathbf{GX}) \right)^H \\ \left(\hat{\mathbf{Y}}-\sqrt{P}\left(\mathbf{FX}+\mathbf{H}\mathrm{diag}(\mathbf{GX})\boldsymbol{\Phi}\right)\right).
\end{multline}
On the other hand, the Riemannian gradient of
$L(\mathbf{X},\boldsymbol{\Phi},\lambda,\rho)$ 
with respect to $\mathbf{X}$ is just the Euclidean gradient
$\nabla_{\mathbf{X}} L$, because $\mathbf{X} \in \mathbb{C}^N$
does not need to be projected.  In this case,
\begin{multline}\label{xgradient}
\nabla_{\mathbf{X}} L  = -2\sqrt{P}\left(\mathbf{H}\mathrm{diag}(\boldsymbol{\Phi}) \mathbf{G}+\mathbf{F}\right)^H \\
\left(\hat{\mathbf{Y}}-\sqrt{P}(\mathbf{H}\mathrm{diag}(\boldsymbol{\Phi}) \mathbf{G}+\mathbf{F})\mathbf{X}\right) \\
\qquad +2(\max(0,{\lambda}+\rho(\|\mathbf{X}\|_2^2-1))\mathbf{X}.
\end{multline}


\begin{algorithm}[t]\label{algo:rcg}
\SetAlgoLined
\SetKw{KwInput}{Input:}
\KwInput{ Starting point $(\mathbf{X}_k,\boldsymbol{\Phi}_k)$, accuracy $\epsilon_k$;
}

Set $j=0$, $\mathbf{X}^{(j)}=\mathbf{X}_k$ and $\boldsymbol{\Phi}^{(j)}=\boldsymbol{\Phi}_k$;


\Repeat 
	{$\|(\nabla_{\mathbf{X}^{(j)}} L,\mathrm{grad}_{\boldsymbol{\Phi}^{(j)}}) L\|_2<\epsilon_k$}
 {

 Compute $\nabla_{\mathbf{X}^{(j)}} L$ and $\mathrm{grad}_{\boldsymbol{\Phi}^{(j)}} L$ using \eqref{xgradient} and \eqref{rgradient};

\eIf{$j=0$}
{
 Set initial search direction $\mathbf{d}_{\mathbf{X}}^{(j)}=-\nabla_{\mathbf{X}^{(j)}} L$ and $\mathbf{d}_{\boldsymbol{\Phi}}^{(j)}=-\mathrm{grad}_{\boldsymbol{\Phi}^{(j)}} L$;
}
{
 Compute $\beta_j$ using Hestenes and Stiefel rule;

 Update the search directions $\mathbf{d}_{\mathbf{X}}^{(j)}$
 and $\mathbf{d}_{\boldsymbol{\Phi}}^{(j)}$ using \eqref{eq:x_search} and
	\eqref{eq:phi_search};
}


 Compute step size $\alpha_j$ by back-tracking line search;

 Update $\mathbf{X}^{(j+1)}$ and $\boldsymbol{\Phi}^{(j+1)}$ using \eqref{eq:rcg_x} and \eqref{eq:rcg_phi};

 Retract $\boldsymbol{\Phi}^{(j+1)}$ using \eqref{retraction};


 $j\leftarrow j+1$

 }

\caption{Riemannian Conjugate Gradient Method for Solving \eqref{alm_sub}}
\end{algorithm}

With the Riemannian gradient defined, the Riemannian conjugate gradient works iteratively as follows. In the $j$-th iteration, the updates of the variables are
\begin{align}
\label{eq:rcg_x}
\mathbf{X}^{(j+1)}&=\mathbf{X}^{(j)}+\alpha_j \mathbf{d}_{\mathbf{X}}^{(j)}, \\
\label{eq:rcg_phi}
\boldsymbol{\Phi}^{(j+1)}&=\boldsymbol{\Phi}^{(j)}+\alpha_j \mathbf{d}_{\boldsymbol{\Phi}}^{(j)},
\end{align}
where $\alpha_j>0$ is the step size found by back-tracking line search on $(\mathbf{X}^{(j)},\boldsymbol{\Phi}^{(j)})$, and $\mathbf{d}_{\mathbf{X}}^{(j)}$ and $\mathbf{d}_{\boldsymbol{\Phi}}^{(j)}$ are the conjugate gradient search directions for $\mathbf{X}$ and $\boldsymbol{\Phi}$, respectively.

For computing the conjugate gradient search direction for $\boldsymbol{\Phi}^{(j+1)}$, since $\mathbf{d}_{\boldsymbol{\Phi}}^{(j)}$ and $\mathbf{d}_{\boldsymbol{\Phi}}^{(j+1)}$ lie in different tangent \mbox{spaces}, i.e., $T_{\boldsymbol{\Phi}^{(j)}}\mathbb{M}_{cc}^N$ and
$T_{\boldsymbol{\Phi}^{(j+1)}}\mathbb{M}_{cc}^N$, respectively, in order to perform the conjugate gradient update rule, a transport, defined as the mapping for a vector from one tangent space to another tangent space is needed.
The transport mapping $\mathcal{T}_{\boldsymbol{\Phi}^{(j)} \rightarrow \boldsymbol{\Phi}^{(j+1)}} :
 T_{\boldsymbol{\Phi}^{(j)}}\mathbb{M}_{cc}^N \mapsto T_{\boldsymbol{\Phi}^{(j+1)}}\mathbb{M}_{cc}^N $ is given by
\begin{multline}
\mathcal{T}_{\boldsymbol{\Phi}^{(j)} \rightarrow \boldsymbol{\Phi}^{(j+1)}}
\left(\mathbf{d}_{\boldsymbol{\Phi}}^{(j)}\right)
= \\
\mathbf{d}_{\boldsymbol{\Phi}}^{(j)} - \Re\left\{\mathbf{d}_{\boldsymbol{\Phi}}^{(j)}\circ \left(\boldsymbol{\Phi}^{(j+1)}\right)^*\right\}\circ\boldsymbol{\Phi}_k^{(j+1)}. 
\end{multline}
With the aid of the transport mapping, the update of $\mathbf{d}_{\boldsymbol{\Phi}}^{(j)}$ is
\begin{equation}\label{eq:phi_search}
\mathbf{d}_{\boldsymbol{\Phi}}^{(j+1)}= -\mathrm{grad}_{\boldsymbol{\Phi}^{(j+1)}} L +\beta_j \mathcal{T}_{\boldsymbol{\Phi}^{(j)} \rightarrow \boldsymbol{\Phi}^{(j+1)}}\left(\mathbf{d}_{\boldsymbol{\Phi}}^{(j)}\right).
\end{equation}
For computing the conjugate gradient search direction for $\mathbf{X}^{(j+1)}$, Euclidean gradient can be directly used and no transport mapping is needed.
Thus, the conjugate gradient update rule is simply
\begin{equation}\label{eq:x_search}
\mathbf{d}_{\mathbf{X}}^{(j+1)}= -\nabla_{\mathbf{X}^{(j+1)}} L +\beta_j \mathbf{d}_{\mathbf{X}}^{(j)}.
\end{equation}
In (\ref{eq:phi_search}) and (\ref{eq:x_search}), the
scalar $\beta_j$ is the conjugate updating parameter which is computed according to the \mbox{Hestenes} and Stiefel rule \cite{cgsurvey2006} on the variables $(\mathbf{X}^{(j)},\boldsymbol{\Phi}^{(j)})$ and $(\mathbf{X}^{(j+1)},\boldsymbol{\Phi}^{(j+1)})$.

After computing the updates (\ref{eq:rcg_x}) and (\ref{eq:rcg_phi}), a final operation called retraction is needed to map the updated vector back onto the manifold. 
Here, $\mathbf{X}^{(j+1)}$ does not need to be retracted.
For $\boldsymbol{\Phi}^{(j+1)}$, the retraction mapping $\mathcal{R}_{\mathbb{M}_{cc}^N} : \mathbb{C}^N \mapsto \mathbb{M}_{cc}^N$
is given as: 
\begin{equation}\label{retraction}
\mathcal{R}_{\mathbb{M}_{cc}^N} \left(\boldsymbol{\Phi}^{(j+1)}\right)
=\mathrm{vec}\left[\frac{\left(\boldsymbol{\Phi}^{(j+1)}\right)_i}
{\left|\left(\boldsymbol{\Phi}^{(j+1)}\right)_i\right|}
\right],
\end{equation}
where $\mathrm{vec}(a_i)$ denotes the operation that takes the elements $a_i, ~i=1,\ldots, N$ to form a column vector.

The overall procedure for solving \eqref{alm_sub} in Algorithm \ref{algo:alm} using the Riemannian conjugate gradient method is summarized as Algorithm \ref{algo:rcg}. The starting point $(\mathbf{X}_k,\boldsymbol{\Phi}_k)$ and the accuracy $\epsilon_k$ come from Algorithm \ref{algo:alm}.

\bibliographystyle{IEEEtran}
\bibliography{SPT}

\newpage
\begin{IEEEbiographynophoto}{Hei Victor Cheng}
	(Member, IEEE) received the B.Eng. degree in electronic engineering from Tsinghua University, Beijing, China, the M.Phil. degree in electronic and computer engineering from the Hong Kong University of Science and Technology, and the Ph.D. degree from the Department of Electrical Engineering, Link\"{o}ping University, Sweden. He was a Post-Doctoral Research Fellow at the University
	of Toronto, Toronto, ON, Canada. He is now an Assistant Professor with the Department of Electrical and Computer Engineering, Aarhus University, Denmark. His current research interests include next generation wireless technologies, intelligent surfaces, and machine learning.
\end{IEEEbiographynophoto}

\begin{IEEEbiographynophoto}{Wei Yu}
	(Fellow, IEEE) received the B.A.Sc. degree in computer engineering and mathematics from the University of Waterloo, Waterloo, ON, Canada, in 1997, and the M.S. and Ph.D. degrees in electrical engineering from Stanford University, Stanford, CA, USA, in 1998 and 2002, respectively. Since 2002, he has been with the Electrical and Computer Engineering Department, University of Toronto, Toronto, ON, Canada, where he is currently a Professor and holds the Canada Research Chair (Tier 1) in Information Theory and Wireless Communications. He is a Fellow of the Canadian Academy of Engineering and a member of the College of New Scholars, Artists, and Scientists of the Royal Society of Canada. Prof. Wei Yu was the President of the IEEE Information Theory Society in 2021, and has served on its Board of Governors since from 2015 to 2023. He served as the Chair of the Signal Processing for Communications and Networking Technical Committee of the IEEE Signal Processing Society from 2017 to 2018. He was an IEEE Communications Society Distinguished Lecturer from 2015 to 2016. Prof. Wei Yu received the Steacie Memorial Fellowship in 2015, the IEEE Marconi Prize Paper Award in Wireless Communications in 2019, the IEEE Communications Society Award for Advances in Communication in 2019, the IEEE Signal Processing Society Best Paper Award in 2008, 2017 and 2021, the Journal of Communications and Networks Best Paper Award in 2017, and the IEEE Communications Society Best Tutorial Paper Award in 2015. He has served as an Area Editor for the IEEE Transactions on Wireless Communications, and as an Associate Editor for IEEE Transactions on Information Theory, IEEE Transactions on Communications, and IEEE Transactions on Wireless Communications.
\end{IEEEbiographynophoto}

\end{document}